\documentclass[sn-mathphys,iicol]{sn-jnl}
\usepackage{amssymb}
\usepackage{amsmath}
\usepackage{amscd}
\usepackage{bbm}
\usepackage{hyperref}
\usepackage{cuted}

\jyear{2022}%

\theoremstyle{thmstyleone}%
\theoremstyle{thmstyletwo}%

\theoremstyle{thmstylethree}%

\title[Dummy header]{Couple stresses and discrete potentials in the vertex model of cellular monolayers}

\author*[1,2]{\fnm{Oliver} E. \sur{Jensen}}\email{oliver.jensen@manchester.ac.uk}

\author[1,2]{\fnm{Christopher} K. \sur{Revell}}\email{christopher.revell@manchester.ac.uk}

\affil[1]{\orgdiv{Department of Mathematics}, \orgname{University of Manchester}, \orgaddress{\street{Oxford Road}, \city{Manchester}, \postcode{M13 9PL}, \country{UK}}}

\affil[2]{\orgdiv{Wellcome Trust Centre for Cell-Matrix Research}, \orgname{University of Manchester}, \orgaddress{\street{Oxford Road}, \city{Manchester}, \postcode{M13 9PL}, \country{UK}}}

\abstract{The vertex model is widely used to simulate the mechanical properties of confluent epithelia and other multicellular tissues.  This inherently discrete framework allows a Cauchy stress to be attributed to each cell, and its symmetric component has been widely reported, at least for planar monolayers.  Here we consider the stress attributed to the neighbourhood of each tricellular junction, evaluating in particular its leading-order antisymmetric component and the associated couple stresses, which characterise the degree to which individual cells experience (and resist) in-plane bending deformations.  We develop discrete potential theory for localised  monolayers having disordered internal structure and use this to derive the analogues of Airy and Mindlin stress functions.  These scalar potentials typically have broad-banded spectra, highlighting the contributions of small-scale defects and boundary-layers to global stress patterns.  An affine approximation attributes couple stresses to pressure differences between cells sharing a trijunction, but simulations indicate an additional role for non-affine deformations.}


\begin{document}

\maketitle

\section{Introduction}

The vertex model is a powerful tool for describing the mechanics of spatially heterogeneous multicellular tissues \cite{weliky1990, nagai2001, farhadifar2007, staple2010, fletcher2014, alt2017}.  A confluent planar epithelium, for example, is represented as polygons tiling a plane.  A mechanical strain energy is attributed to each cell that is a function of geometric invariants (such as the cell's area and perimeter) and the total energy of the monolayer is minimised, at a rate defined via a model of viscous dissipation, by varying vertex locations, potentially allowing for cell neighbour exchanges (so-called T1 transitions).   A force balance at each vertex is used to evolve the system to equilibrium; elastic forces are defined by taking the first variation of each cell's mechanical energy with respect to vertex displacements.  The changes of a cell's area and perimeter arising from small displacements of its vertices can thereby be used to define the mechanical (Cauchy) stress attributed to each cell.  The model predicts a symmetric Cauchy stress tensor associated with each cell \cite{ishihara2012, yang2017} that aligns with cell shape \cite{ANB2018a} and allows viscoelastic moduli for bulk and shear deformations to be evaluated \cite{ANB2018b, tong2021}.   Less attention has been paid to the Cauchy stress defined over the network that is topologically dual to cellular polygons, namely the triangulation connecting adjacent cell centres.   The stress attributed to each triangle describes the mechanical environment in the neighbourhood of the tricellular junction lying within the triangle.  This stress field is of interest given the role of tricellular junctions as potential sensors of cell shape and mechanical stress \cite{higashi2017, bosveld2018, nestor2019, angulo2020, yu2020}.

From a multiscale modelling perspective, the vertex model is also of interest as a bridge between descriptions of discrete cells in a tissue and a continuum description of the tissue's mechanical properties \cite{murisic2015, tlili2015, ishihara2017}.  In two-dimensional (2D) continuum mechanics, it is often convenient to express the Cauchy stress in terms of a scalar potential, the Airy stress function \cite{howell2009}.  However in seeking to construct the discrete analogue of the Airy stress function, we found \cite{jensen2020} that the requirement for both forms of the Cauchy stress (that defined over cells, and that defined over tricellular junctions) to be symmetric places severe geometric constraints on cell shape, specifically that cell edges should be orthogonal to links between cell centres and that each vertex should lie at the orthocentre of the triangle formed by its immediate neighbours. These constraints are not met in typical simulations (nor, indeed, in real monolayers).  This discrepancy can be explained in part by noting that while forces balance at vertices in the normal implementation of the vertex model, torque balance is not enforced.  Here we consider how the discrepancy can be accommodated by relaxing the requirement for all Cauchy stresses to be symmetric, by incorporating couple stresses within the constitutive framework.  This approach is natural given the use of second-gradient or micropolar models to describe materials with microstructure \cite{trovalusci2015, rizzi2019}.

The Cauchy stress attributed to a cell (which hereafter we call the force stress, evaluated as the first spatial moment of the forces acting over a cell) can be partitioned into an isotropic component (defining an effective cell pressure) and a deviatoric component (describing the shear stress experienced by each cell) \cite{ANB2018a}.  Analogous quantities can be attributed to the triangles bounding tricellular junctions, having vertices at cell centres.  Couple stress provides an additional measure of the stress arising from in-plane bending deformations {\color{black}that generate curvature in material elements}.  Using a standard version of the vertex model, we demonstrate here that while individual cells experience zero torque, a couple can be exerted around tricellular junctions.  By considering second-order spatial gradients of a virtual tissue deformation, we show how the couple can be explained, in part, by considering the degree to which a cell is `bent out of shape' via pressure differences creating moments acting across adjacent cell edges.  However, our analysis reveals limits to potential analogies between the vertex model and continuum theories of couple-stress materials, likely associated with non-affine deformations occuring at small scales.

Our calculations are facilitated through the use of tools of discrete calculus \cite{grady2010}.  In particular, incidence matrices capture topological relationships between cell vertices, edges and faces and enable the primal network of polygonal cells to be related directly to the dual triangulation connecting adjacent cell centres.  Incidence matrices also provide the building blocks of the discrete differential operators needed to represent stresses using vector and scalar potentials.  Unlike the three operators needed for normal continuum mechanics (grad, div and curl), we find that up to 16 different operators (4 grads, 4 divs and 8 curls) are required in two spatial dimensions, in the general instance when links between cell centres are not orthogonal to cell edges.  These operators permit representations of spatially-2D vectors in terms of scalar potentials, via Helmholtz--Hodge decomposition.  For 2D continuum elasticity, two potentials suffice for simply-connected domains (the Airy stress function, plus an additional stress function defined by Mindlin for couple-stress materials \cite{mindlin1962, hadjesfandiari2011}).  For discrete networks of cells, we find that up to eight potentials typically emerge, four defined over the network of cells, and four over the dual triangulation, although these reduce in number when edges and links are orthogonal.  The potentials facilitate visualisation of stress patterns across a monolayer and their construction in terms of eigenmodes of scalar Laplacians, built using the geometry and topology of the cell network, reveals how stress fields are influenced both by the macroscopic shape of a localised monolayer and small-scale features such as topological defects in the organisation of individual cells.

We briefly review continuous couple-stress materials in 2D in Appendix~\ref{sec:couple}, following \cite{hadjesfandiari2011}.  Key points to highlight are: (i) the Cauchy stress $\boldsymbol{\sigma}$ and couple stress vector $\boldsymbol{\mu}$ can be written in terms of continuous Airy and Mindlin stress functions $\psi$ and $\Psi$ in the the form
\begin{equation}
\boldsymbol{\sigma} =\mathrm{curl}\otimes (\mathrm{curl}\, \psi - \mathrm{grad}\, \Psi), \quad \boldsymbol{\mu}=-\mathrm{curl}\, \Psi,
\label{eq:curlcurl}
\end{equation}
thus satisfying force balance $\mathrm{div}\,\boldsymbol{\sigma}=\mathbf{0}$, a torque balance relating the antisymmetric component of $\boldsymbol{\sigma}$ to $\mathrm{curl}\,\boldsymbol{\mu}$, and a compatibility condition (derived from a constitutive assumption) $\mathrm{div}\,\boldsymbol{\mu}=0$; (ii) the vector potential $\mathrm{curl}\, \psi - \mathrm{grad}\, \Psi$ is here expressed using a Helmholtz decomposition in terms of the two scalar potentials $\psi$ and $\Psi$; and (iii) in the principle of virtual work, the strain $\tfrac{1}{2}(\nabla \mathbf{u}+\nabla \mathbf{u}^\top)$ of a small-amplitude deformation $\mathbf{u}(\mathbf{x})$ is energy-conjugate to $\boldsymbol{\sigma}$ while $\tfrac{1}{2} (\nabla^2 \mathbf{u}-\nabla(\nabla\cdot \mathbf{u}))=-2\boldsymbol{\kappa}$ (a strain gradient), where $\boldsymbol{\kappa}$ is the so-called curvature, is energy-conjugate to $\boldsymbol{\mu}$.    We seek discrete analogues of these relationships below (disregarding the compatibility condition $\mathrm{div}\,\boldsymbol{\mu}=0$, as we make an alternative constitutive assumption), starting by describing the nature of Helmholtz decomposition over a discrete cellular network and its dual triangulation.  Key aspects of discrete calculus that we exploit are summarised in Sec.~\ref{sec:model}, with details provided in Appendix~\ref{sec:operators}.  In particular, we identify four Laplacians associated with the 16 operators, through which scalar potentials can (in principle) be derived to describe any vector field defined over the cell network.  The eigenmodes of the Laplacians, which are shaped by the boundary of the monolayer and the organisation of individual cells within it, provide the building blocks for stress fields.  In Sec.~\ref{sec:3}, we show how stresses in a vertex model can be expressed in terms of a force potential (following \cite{jensen2020}) and determine the underlying scalar potentials.  Using a standard constitutive model, we evaluate in Sec.~\ref{sec:const} the couple stress.  Here the analogy between discrete and continuous descriptions is revealing but imperfect, as the couple stress determined as a rotational component of a vector force potential (as in (\ref{eq:curlcurl})) turns out to differ from the vector that is energy conjugate to $\boldsymbol{\kappa}$ (under an affine approximation), for the particular constitutive model that we investigate.  Results are illustrated by computations in Sec.~\ref{sec:comp} and discussed in Sec.~\ref{sec:disc}.

\section{Discrete cellular calculus}
\label{sec:model}

Before addressing mechanical questions in Sec.~\ref{sec:3}, it is necessary to develop relevant tools of calculus for quantities defined over a disordered cellular monolayer.  We use topological (Sec.~\ref{sec:topol}) and geometric (Sec,~\ref{sec:geom}) objects to derive operators (Sec.~\ref{sec:ops}), in particular discrete Laplacians, enabling vectors to be represented in terms of scalar potentials (Sec.~\ref{sec:helm}) built from eigenmodes of Laplacians that are specific to the monolayer (Sec.~\ref{sec:potthy}).

\subsection{Cell topology}
\label{sec:topol}

\begin{figure*}
    \centering
    \includegraphics[width=\textwidth]{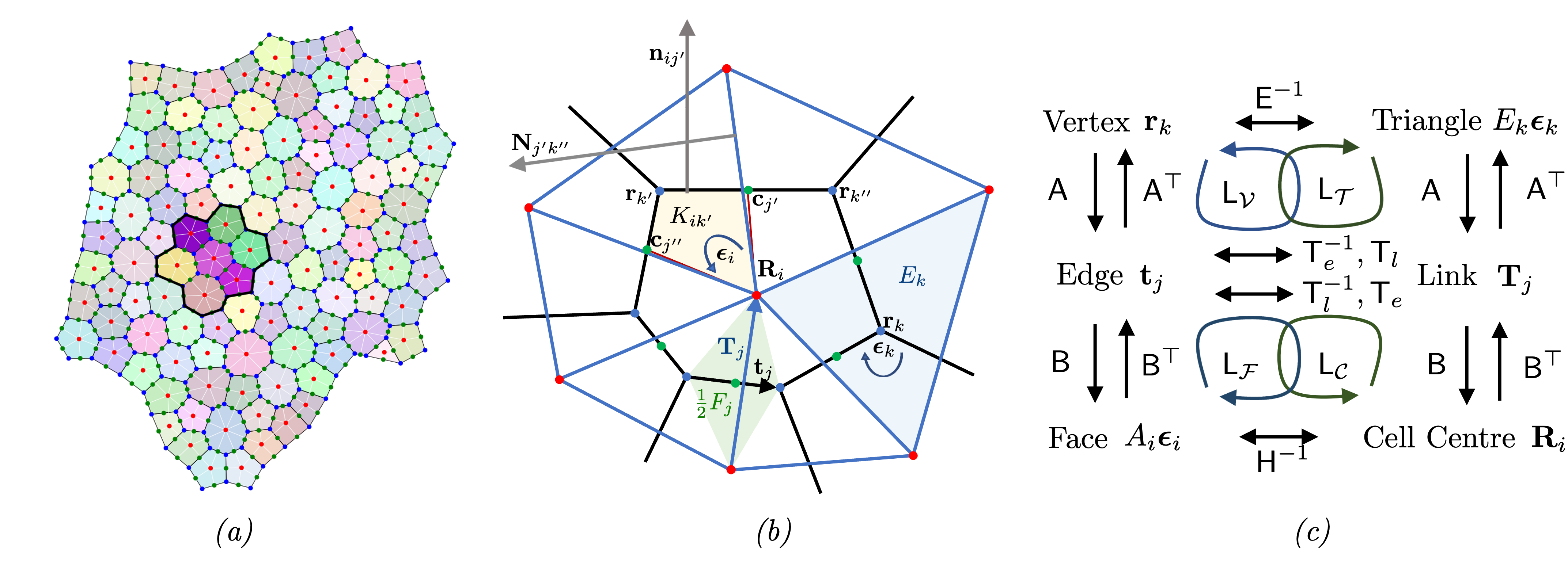}
    \caption{
    (a) A monolayer of $N_c=112$ cells, with $N_e=345$ edges and $N_v=234$ vertices, `grown' using the vertex model via a sequence of random cell divisions.  The primal network is defined by cell vertices (blue dots) and edges (black lines); the dual network is defined by cell centres (red dots) and links (white lines); white links also connect centres of border cells to  peripheral edge centroids (green dots).  A cluster of 7 cells, used in Figure~\ref{fig:forcenet}(a) below, is highlighted {\color{black}with bolder colours}. 
    (b) A sketch defining geometric objects, their orientations and labels.  Black lines denote cell edges, passing through vertices (blue dots) including $\mathbf{r}_k$, $\mathbf{r}_{k'}$ and $\mathbf{r}_{k''}$; blue lines denote links between cell centres (red dots), including $\mathbf{R}_i$.  Yellow: the kite of cell $i$ at vertex $k'$ with area $K_{ik'}$, with two of its vertices at edge centroids (green dots) $\mathbf{c}_j$ and $\mathbf{c}_{j''}$. Green: the trapezium with area $\tfrac{1}{2} F_j$ spanned by edge $\mathbf{t}_j$ and link $\mathbf{T}_j$ (orientations of other edges and links are not shown). Blue: the triangle surrounding vertex $k$ with area $E_k$.  Also shown are the outward normals $\mathbf{n}_{ij'}$ to cell $i$ at edge $j'$, and $\mathbf{N}_{j'k''}$, to triangle $k''$ at edge $j'$. Cell orientations $\boldsymbol{\epsilon}_i$ are opposite to triangle orientations $\boldsymbol{\epsilon}_k$.  
    (c) A diagram indicating how incidence matrices $\mathsf{A}$ and $\mathsf{B}$ map between vertices, edges, faces on the primal network, and cell centres, links and triangles on the dual network.  Metric matrices $\mathsf{E}^{-1}$, $\mathsf{T}_e$, $\mathsf{T}_l$ and $\mathsf{H}^{-1}$ map between networks; {\color{black} $\mathsf{T}_e$ and $\mathsf{T}_e^{-1}$ map from edges to links and $\mathsf{T}_l$ and $\mathsf{T}_l^{-1}$ from links to edges}.  Loops indicate how the Laplacians $\mathsf{L}_{\mathcal{V}}$, $\mathsf{L}_{\mathcal{T}}$, $\mathsf{L}_{\mathcal{C}}$, $\mathsf{L}_{\mathcal{F}}$ in (\ref{eq:lapmat}) are constructed.}
    \label{fig:geometry}
\end{figure*}

We consider an isolated cellular monolayer occupying a simply-connected domain on the Euclidean plane, as illustrated in Figure~\ref{fig:geometry}(a).  Adopting notation used in \cite{jensen2020}, vertices, edges and faces of the (primal) cell network are labelled by $k$, $j$ and $i$ respectively (Figure \ref{fig:geometry}(b)), where $i=1,\dots,N_c$, $j=1,\dots,N_e$ and $k=1,\dots,N_v$.  Orientations are assigned to each object, and the topological relationships between edges and vertices, and faces and edges, are defined by the signed incidence matrices $A_{jk}$ and $B_{ij}$ respectively.  (Thus $A_{jk}=1$ if edge $j$ points into vertex $k$, $A_{jk}=-1$ if edge $j$ points out of vertex $k$, and $A_{jk}=0$ otherwise; $B_{ij}=1$ if edge $j$ neighbours cell $i$ and has congruent orientation, $B_{ij}=-1$ if edge $j$ neighbours cell $i$ but has opposite orientation, and $B_{ij}=0$ otherwise.)  $\mathsf{A}$ and $\mathsf{B}$ also specify topological relationships between cell centres (assumed here to be cell vertex centroids), links connecting cell centres, and triangular faces of the dual network (Figure~\ref{fig:geometry}c).  Vertices within the interior of a monolayer are assumed to neighbour three cells: vertex/face neighbours are identified by the unsigned adjacency matrix $\mathsf{C}=\tfrac{1}{2}\overline{\mathsf{B}} \,\overline{\mathsf{A}}$, where $\overline{\mathsf{A}}$ and $\overline{\mathsf{B}}$ are unsigned incidence matrices (where $\overline{A}_{jk}\equiv \vert A_{jk}\vert$, $\overline{B}_{ij}\equiv \vert B_{ij} \vert$).   Neighbour exchanges (leading to plastic deformations, with consequences for cell packing \cite{hashimoto2018}) are incorporated in simulations but otherwise not considered in the present study, so that incidence matrices remain fixed.  The topological identity $\mathsf{B}\mathsf{A}=\mathsf{0}$ (which can be interpreted by saying that the boundary of any localised clump of cells is closed and therefore has no boundary) underpins the construction of discrete differential operators.   We also identify centroids of each edge: each cell can then be partitioned into kites (labelled by $ik$, see Figure~\ref{fig:geometry}b).   
In general, links between cell centroids in this network do not pass through edge centroids, except for those in cells at the periphery of the monolayer.  Thus the faces of the dual network are internal triangles, or kites within cells at the periphery of the monolayer (either a single kite, or a pair of kites in adjacent cells).

\subsection{Cell geometry}
\label{sec:geom}

We introduce geometric information as follows.  Points in the underlying Euclidean plane have position vector $\mathbf{x}$.   Where necessary, $p$, $q$, $r$ denote subscripts of vectors and tensors, identifying components with respect to a fixed basis in this plane, and a bold font is used to denote vectors in $\mathbbm{R}^2$.  On the primal network of cells, we define vertices by $\mathbf{r}_k$, edges by $\mathbf{t}_j=\sum_k A_{jk}\mathbf{r}_k$, edge lengths by $t_j=\vert\mathbf{t}_j\vert$ and edge centroids by $\mathbf{c}_j=\tfrac{1}{2}\sum_k \overline{A}_{jk} \mathbf{r}_k$.  As indicated in Figure~\ref{fig:geometry}(c), $\mathsf{A}$ acts as a difference operator in mapping vertex locations $\mathbf{r}_k$ to edges $\mathbf{t}_j$.  In contrast, $\mathsf{A}^\top$ and $\mathsf{B}^\top$ act as boundary operators.  For example, the non-zero elements of $\mathbf{c}^p_j= \sum_i \vert \mathbbm{1}^c_i {B}_{ij} \vert  \mathbf{c}_j$ are the edge centroids around the periphery of the monolayer, where $\mathbbm{1}^c\equiv (1,1,\dots, 1)$ is the $N_c$-vector (a 2-chain, in the language of discrete calculus) denoting all cells in the monolayer.  The number of edges of cell $i$ is given by $Z_i=\sum_j \overline{B}_{ij}$.  We define the centre of cell $i$ as $\mathbf{R}_i=Z_i^{-1} \sum C_{ik} \mathbf{r}_k$.  Links on the dual network, triangulating cell centres, are defined by
\begin{equation}
\mathbf{T}_j={\textstyle\sum_i} B_{ij}(\mathbf{R}_i-\mathbf{c}^p_j),
\label{eq:links}
\end{equation}
so that links either connect cell centres or connect centres of border cells (at the periphery of the monolayer) to peripheral edge centroids.  Orientations of cell faces on the primal network $\boldsymbol{\epsilon}_i$, and triangles (or peripheral kites) on the dual network $\boldsymbol{\epsilon}_k$, are prescribed as $\pm \boldsymbol{\varepsilon}$, where the matrix $\boldsymbol{\varepsilon}$ (the 2D Levi-Civita tensor) represents a clockwise $\pi/2$ rotation.  $\boldsymbol{\epsilon}_i$ and $\boldsymbol{\epsilon}_k$ are taken to be independent of $i$ and $k$ respectively and of opposite sense.  To ensure that $\mathsf{A}$ and $\mathsf{B}$ apply also to the dual network, orientations of edges $\mathbf{t}_j$ and links $\mathbf{T}_j$ are constrained such that
\begin{equation}
\label{eq:trapezium}
\mathbf{T}_j\cdot \boldsymbol{\epsilon}_i\mathbf{t}_j =\mathbf{t}_j \cdot \boldsymbol{\epsilon}_k \mathbf{T}_j = F_j>0,
\end{equation}
where $\tfrac{1}{2}F_j$ is the area of the trapezium spanned by $\mathbf{t}_j$ and $\mathbf{T}_j$ (for interior edges, Fig.~\ref{fig:geometry}b) or the area of the triangle spanned by $\mathbf{t}_j$ and $\mathbf{T}_j$ (for peripheral edges).  Consistent with typical simulations of the vertex model \cite{jensen2020}, we allow edges and links to be non-orthogonal.

For cell $i$, the outward normal to edge $j$ is $\mathbf{n}_{ij}=-\boldsymbol{\epsilon}_i B_{ij} \mathbf{t}_j$.  Likewise the outward normal to the triangle connecting adjacent cell centres is $\mathbf{N}_{jk}=-\boldsymbol{\epsilon}_k A_{jk} \mathbf{T}_j$ (Figure~\ref{fig:geometry}b).  The areas of cells and of interior triangles, $A_i$ and $E_k$, satisfy
\begin{equation}
{\textstyle \sum_j} B_{ij} \mathbf{t}_j\otimes \mathbf{c}_j=A_i \boldsymbol{\epsilon}_i, \quad
{\textstyle \sum_j} A_{jk} \mathbf{T}_j\otimes \mathbf{C}_j=E_k \boldsymbol{\epsilon}_k,
\label{eq:cellarea}
\end{equation}
where centroids of internal links are defined by $\mathbf{C}_j=\tfrac{1}{2}\sum_i \overline{B}_{ij} \mathbf{R}_i$.  Thus $A_i = \tfrac{1}{2} \sum_j \mathbf{n}_{ij}\cdot \mathbf{c}_j$, and the cell perimeter satisfies $L_i=\sum_j \overline{B}_{ij} t_j$.  Similarly $E_k = \tfrac{1}{2} \sum_j \mathbf{N}_{jk}\cdot \mathbf{C}_j$ gives the area of each triangle at interior vertex $k$; $E_k$ is defined as the area of the adjacent kite (or kites) if $k$ identifies a peripheral vertex (Figure~\ref{fig:geometry}a).   The total monolayer area $\mathcal{A}$ can then be written
\begin{equation}
\mathcal{A}={\textstyle{ \sum_i}} A_i = {\textstyle{ \sum_j}} \tfrac{1}{2} F_j = {\textstyle \sum_k} E_k,
\end{equation}
indicating how the monolayer can be partitioned into cells (labelled by $i$), trapezia spanned by edges and links (labelled by $j$) and triangles (plus peripheral kites,  labelled by $k$).


To summarise, all topological information is encoded in $\mathsf{A}$ and $\mathsf{B}$, while metric information is encoded in edge and link lengths $t_j$, $T_j=\vert \mathbf{T}_j \vert$ and in the areas $A_i$, $E_k$ and $F_j$.  Using these we define the matrices
$\mathsf{H}=\mathrm{diag}(A_1,\dots,A_{N_c})$, $\mathsf{E}=\mathrm{diag}(E_1,\dots,E_{N_v} )$,
$\mathsf{T}_e=\mathrm{diag}\left({t_1^2}/{F_1},\dots, {t_{N_e}^2}/{F_{N_e}}\right)$,
$\mathsf{T}_l=\mathrm{diag} \left({T_1^2}/{F_1},\dots ,{T_{N_e}^2 }/{F_{N_e} }\right)$
with which we can define the square matrices
\begin{subequations}
\label{eq:lapmat}
\begin{gather}
\mathsf{L}_{\mathcal{V}}=\mathsf{E}^{-1} \mathsf{A}^\top \mathsf{T}_e^{-1} \mathsf{A}, \quad
\mathsf{L}_{\mathcal{T}}=\mathsf{E}^{-1} \mathsf{A}^\top \mathsf{T}_l \mathsf{A}, \\
\mathsf{L}_{\mathcal{F}}=\mathsf{H}^{-1} \mathsf{B} \mathsf{I}_b \mathsf{T}_e \mathsf{B}^\top, \quad
\mathsf{L}_{\mathcal{C}}=\mathsf{H}^{-1} \mathsf{B} \mathsf{I}_b \mathsf{T}_l^{-1} \mathsf{B}^\top.
\end{gather}
\end{subequations}
(The diagonal matrix $\mathsf{I}_b\equiv \mathrm{diag}(\mathbbm{1}^e-\sum_i \vert \mathbbm{1}^c_i B_{ij}\vert)$ eliminates `orphan' links at the monolayer periphery that connect to only one cell centre.)  The construction of these operators, which turn out to be the Laplacians for scalar fields defined over vertices, triangles, cell centres or cell faces respectively {\color{black}(Appendix~\ref{sec:operators})}, is illustrated in Figure~\ref{fig:geometry}(c) and explained in more detail below.

\subsection{Discrete operators}
\label{sec:ops}

\begin{figure*}
    \centering
    \includegraphics[width=0.8\textwidth]{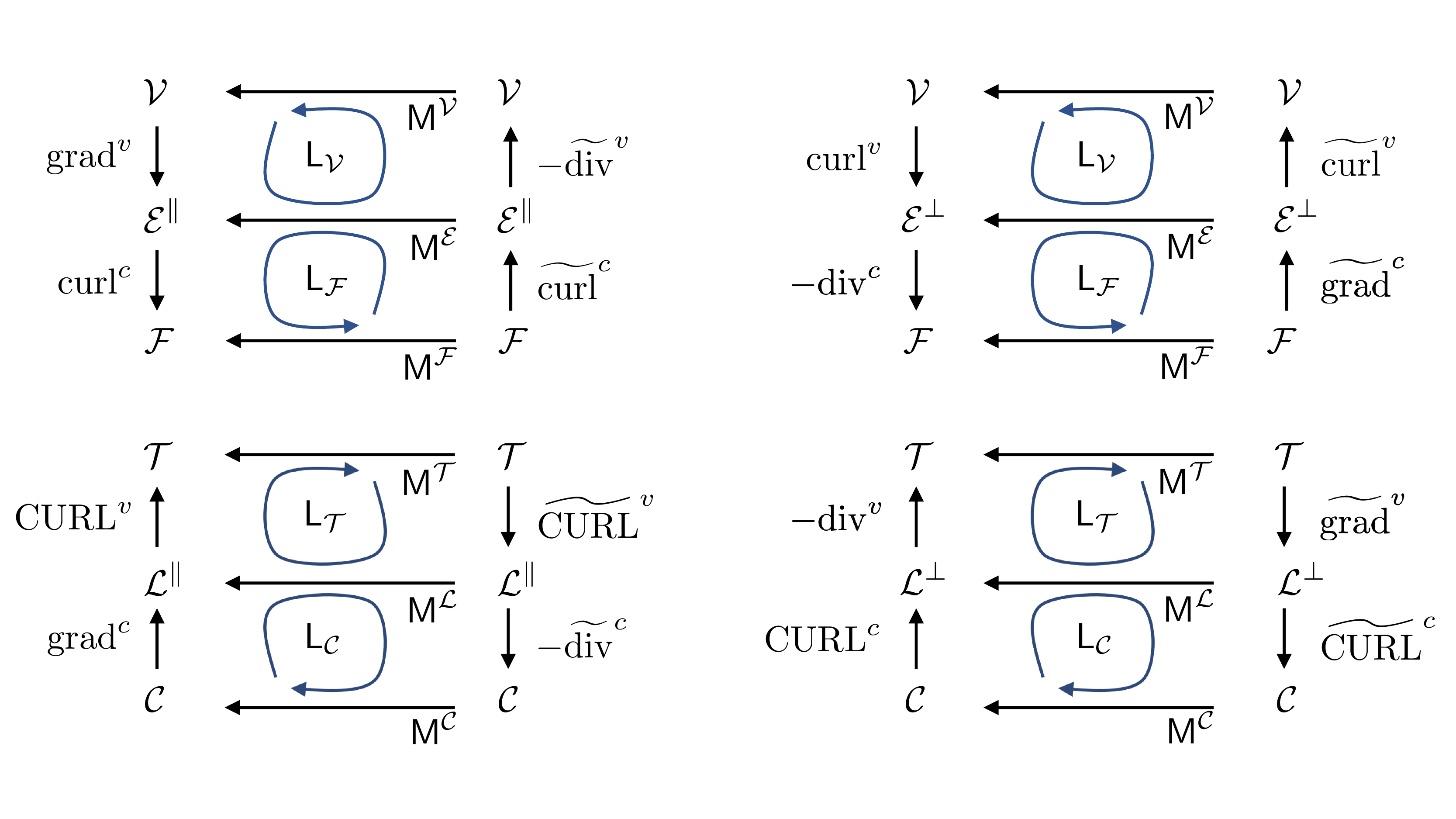}
    \caption{Four diagrams showing the action of operators defined on the primal network of cells (top) and dual network of triangles (bottom), involving vectors parallel (left) and perpendicular (right) to edges and links respectively.  In each diagram, primary operators run along left-hand vertical arrows.  Derived operators, running along right-hand vertical arrows, are adjoint to primary operators under inner products (horizontal arrows) acting on elements of vector spaces $\mathcal{V}\simeq \mathcal{T}$ (scalar functions defined on vertices and triangles), $\mathcal{E}\simeq \mathcal{L}$ (vectors defined on edges and links), $\mathcal{F}\simeq \mathcal{C}$ (scalar functions defined on cell faces and centres), where $\simeq$ denotes isomorphism.  Loops show how operators and inner products create scalar Laplacians $\mathsf{L}_\mathcal{V}$, $\mathsf{L}_\mathcal{F}$, $\mathsf{L}_\mathcal{T}$, $\mathsf{L}_\mathcal{C}$.  Adjacent pairs of vertical arrows indicate exact sequences, such as $\mathrm{div}^v\circ \mathrm{CURL}^c=0$.  Inner products are evaluated using matrices $\mathsf{M}^\mathcal{V}=\mathsf{M}^\mathcal{T}$, $\mathsf{M}^\mathcal{E}=\mathsf{M}^\mathcal{L}$, $\mathsf{M}^\mathcal{F}=\mathsf{M}^\mathcal{C}$, defined in Appendix~\ref{sec:operators}.}
    \label{fig:operators}
\end{figure*}

Variables (so-called co-chains) defined over cells, edges or vertices are written without the $i$, $j$, $k$ subscript, so that $\{A \}_i=A_i$, $\{ \mathbf{c} \}_j=\mathbf{c}_j$, $\{ E \}_k=E_k$, etc. The scalar fields $A$, $F$ and $E$ are used below to define inner products with which discrete differential operators are constructed.  In  Appendix~\ref{sec:operators}, we describe how discrete analogues of grad, div and curl operators for scalar-valued variables defined on vertices or cell centres, and vector-valued variables defined on edges or links, can be defined.   Figure~\ref{fig:operators} illustrates how the 16 operators act.  Explicit expressions for the 8 so-called primary operators are given in Table~\ref{table1}.  To summarise briefly, vector-valued functions defined on edges or links sit in the isomorphic vector spaces $\mathcal{E}$ and $\mathcal{L}$ respectively, which can be partitioned into subspaces $\mathcal{E}=\mathcal{E}^\parallel\oplus\mathcal{E}^\perp$ (or $\mathcal{L}=\mathcal{L}^\parallel\oplus\mathcal{L}^\perp$) of vectors parallel and pependicular to edges (or links).  Gradient operators $\mathrm{grad}^v$ and $\mathrm{grad}^c$ act on scalars defined at vertices and cell centres (in vector spaces $\mathcal{V}$ and $\mathcal{C}$ respectively), creating vectors in $\mathcal{E}^\parallel$ and $\mathcal{L}^\parallel$ respectively.  Curl operators acting on scalars, $\mathrm{curl}^v$ and $\mathrm{CURL}^c$, are rotated gradients that create vectors that are normal to edges and links respectively (in $\mathcal{E}^\perp$ and $\mathcal{L}^\perp$).  Divergence operators $\mathrm{div}^v$ and $\mathrm{div}^c$ measure fluxes of vectors normal to edges and links, mapping vectors from $\mathcal{E}^\perp$ and $\mathcal{L}^\perp$ to scalars defined over faces and triangles (in spaces $\mathcal{F}$ and $\mathcal{T}$ respectively).  Curl operators acting on vectors, $\mathrm{CURL}^v$ and $\mathrm{curl}^c$, are similar, but measure fluxes parallel to edges and links, mapping from $\mathcal{E}^\parallel$ and $\mathcal{L}^\parallel$ to $\mathcal{F}$ and $\mathcal{T}$ respectively. Via the fundamental relationship $\mathsf{B}\mathsf{A}=\mathsf{0}$, these operators respect the exact relationships $\mathrm{curl}^c\circ\, \mathrm{grad}^v=0$, $\mathrm{div}^c \,\circ\,\mathrm{curl}^v=0$ and so on, as summarised in Fig.~\ref{fig:operators}.  Superscripts $v$ and $c$ are used to denote objects associated with cells and vertices respectively, and therefore primarily involve $\mathsf{B}$ and $\mathsf{A}$ respectively.

\begin{table*}
    \centering
    \begin{tabular}{| c | c || c | c |}
        \hline
        $\{ \mathrm{grad}^v\,\phi \}_j$ & ${\textstyle \sum_k} A_{jk}({\mathbf{t}_j}/{t_j^2})\phi_k$ &
        $\{ \mathrm{curl}^v \phi \}_j$ & ${\textstyle \sum_k} \boldsymbol{\epsilon}_k (\mathbf{t}_j/{t_j^2} )A_{jk} \phi_k $ \\
         $\{ \mathrm{curl}^c\,\mathbf{b}\}_i $ & ${\textstyle \sum_j} B_{ij} \mathbf{t}_j \cdot \mathbf{b}_j /A_i $  &
        $\{ \mathrm{div}^c\,\mathbf{b}\}_i$ & $- {\textstyle \sum_j} B_{ij} (\boldsymbol{\epsilon}_i \mathbf{t}_{j}) \cdot \mathbf{b}_j /A_i$ \\
         $\{ \mathrm{grad}^c\,f \}_j $ & ${\textstyle \sum_i} B_{ij}({\mathbf{T}_j}/{T_j^2}) f_i$ &
        $\{ \mathrm{CURL}^c f \}_j $ & ${\textstyle \sum_i}  \boldsymbol{\epsilon}_i (\mathbf{T}_j /{T_j^2} )B_{ij} f_i$ \\
         $\{ \mathrm{CURL}^v\,\mathbf{b}\}_k $ & $ {\textstyle \sum_j} A_{jk} \mathbf{T}_j \cdot \mathbf{b}_j / E_k $  &
        $\{ \mathrm{div}^v\,\mathbf{b}\}_k$ &
        $- {\textstyle \sum_j}  A_{jk} (\boldsymbol{\epsilon}_k \mathbf{T}_{j}) \cdot \mathbf{b}_j /E_k $  \\ [0.5ex]
        \hline \hline
                $\{ \widetilde{\mathrm{CURL}}^v\, \phi \}_j $ & $ {\textstyle \sum_k} A_{jk} \mathbf{T}_j \phi_k/F_j $
        &         $\{ \widetilde{\mathrm{grad}}^v\, \phi \}_j $ & ${\textstyle \sum_k} A_{jk} \boldsymbol{\epsilon}_k (\mathbf{T}_j /F_j)\phi_k $     \\
              $\{ \widetilde{\mathrm{div}}^c\,\mathbf{b}\}_i$ & $- {\textstyle \sum_j} B_{ij} ({F_j}/ {T_j^2}) \mathbf{T}_j\cdot \mathbf{b}_j/A_i $
        &         $\{ \widetilde{\mathrm{CURL}}^c \mathbf{b}  \}_i $ & $ {\textstyle \sum_j} B_{ij} ({F_j}/{T_j}^2) (\boldsymbol{\epsilon}_i\mathbf{T}_j) \cdot \mathbf{b}_j/A_i $ \\
         $\{ \widetilde{\mathrm{curl}}^c \,f \}_j $ & $ {\textstyle \sum_i} B_{ij} (\mathbf{t}_j /{F_j})  f_i$         &
        $\{ \widetilde{\mathrm{grad}}^c \,f \}_j $ & ${\textstyle \sum_i} B_{ij} \boldsymbol{\epsilon}_i (\mathbf{t}_j / F_j) f_i$         \\
        $ \{ \widetilde{\mathrm{div}}^v \,\mathbf{b} \}_k $ & $- {\textstyle \sum_j} A_{jk} ({F_j}/{t_j^2}) \mathbf{t}_j \cdot \mathbf{b}_j /E_k$
        & $\{ \widetilde{\mathrm{curl}}^v \,\mathbf{b}\}_k $ & $ {\textstyle \sum_j}A_{jk} ({F_j}/{t_j^2}) (\boldsymbol{\epsilon}_k \mathbf{t}_j)\cdot \mathbf{b}_j/E_k$
        \\
        \hline
    \end{tabular}
    \caption{A summary of the 16 discrete operators.   The top eight are primary, the lower eight are derived.  Those on the left-hand (right-hand) half of the table map to, or act on, vectors parallel (perpendicular) to edges or links respectively.  Alternative expressions for $\mathrm{CURL}^v$, $\mathrm{div}^v$, $\widetilde{\mathrm{CURL}}^v$ and $\widetilde{\mathrm{grad}}$ that apply also to peripheral kites are given in (\ref{eq:periphop}).}
    \label{table1}
\end{table*}


When edges and links are non-orthogonal ($\mathbf{t}_j\cdot \mathbf{T}_j\neq 0$), as we assume to be the case, a further eight so-called derived operators (adjoints under a suitable inner product, denoted with a tilde) must be considered (illustrated in Fig.~\ref{fig:operators}); definitions are given in Table~\ref{table1}.  Thereby we derive scalar Laplacians $\mathsf{L}_{\mathcal{F}}=-{\mathrm{div}}^c \,\circ \, \widetilde{\mathrm{grad}}^c={\mathrm{curl}}^c\,\circ\,\widetilde{\mathrm{curl}}^c$ acting on cell faces, and $\mathsf{L}_\mathcal{V}=-\widetilde{\mathrm{div}}^v \,\circ \, \mathrm{grad}^v=\widetilde{\mathrm{curl}}^v\,\circ\,\mathrm{curl}^v$ on vertices,
given in matrix form by 
(\ref{eq:lapmat}a).
On the dual network, these have analogues
$\mathsf{L}_{\mathcal{C}}=-\widetilde{\mathrm{div}}^v \,\circ \, {\mathrm{grad}}^c={\widetilde{\mathrm{CURL}}}^c\,\circ\,{\mathrm{CURL}}^c$ acting on cell centres and $\mathsf{L}_\mathcal{T}=-{\mathrm{div}}^v \,\circ \, \widetilde{\mathrm{grad}}^v={\mathrm{CURL}}^v\,\circ\,\widetilde{\mathrm{CURL}}^v$ on triangles, given by (\ref{eq:lapmat}b).  The four scalar Laplacians reduce to two ($\mathsf{L}_{\mathcal{T}}=\mathsf{L}_{\mathcal{V}}$, $\mathsf{L}_{\mathcal{F}}=\mathsf{L}_{\mathcal{C}}$) in the special case of edge-link orthogonality, when $F_j=T_j t_j$.



\subsection{Helmholtz decomposition}
\label{sec:helm}

A vector defined over edges or links can be represented in terms of potentials defined over each network, via a form of Helmholtz decomposition.  (We do not provide a formal proof but call on analogous results developed for mimetic finite differences \cite{daveiga2014}.)  Assuming the cell monolayer is simply connected, any vector $\mathbf{h}\in\mathcal{E}$ with elements $\mathbf{h}_j$ ($j=1,\dots,N_e$) has representation over the primal network of the form $\mathbf{h}=\mathbf{h}^\parallel+\mathbf{h}^\perp$, where
\begin{subequations}
    \label{eq:h1}
\begin{align}
    \mathbf{h}^\parallel &=\mathrm{grad}^v\psi^v + \widetilde{\mathrm{curl}}^c \Psi^c \in \mathcal{E}^\parallel, \\ 
    \mathbf{h}^\perp &=\widetilde{\mathrm{grad}}^c\psi^c + {\mathrm{curl}}^v \Psi^v \in \mathcal{E}^\perp,
\end{align}
\end{subequations}
%
for some $\psi^v$ and $\Psi^v \in \mathcal{V}$ (with components $\psi^v_k$, $\Psi^v_k$, $k=1,\dots, N_v$), and for some $\psi^c$ and $\Psi^c\in\mathcal{F}$ (with components $\psi^c_i$, $\Psi^c_i$, $i=1,\dots,N_c$).  In (\ref{eq:h1}), $\mathbf{h}$ has been decomposed into its components parallel and perpendicular to each edge. 
Noting that $-\mathrm{div}^c\,\mathbf{h} =-\mathrm{div}^c \,\mathbf{h}^\perp =-\mathrm{div}^c\,\circ\, \widetilde{\mathrm{grad}}^c \psi^c=\mathsf{L}_\mathcal{F}\psi^c$ (and so on), we see that the potentials are determined by solving the Poisson problems
\begin{subequations}
\label{eq:pot1}
\begin{gather}
\mathsf{L}_{\mathcal{F}} \psi^c=-\mathrm{div}^c \,\mathbf{h},
\quad
\mathsf{L}_{\mathcal{F}} \Psi^c = \mathrm{curl}^c\,\mathbf{h},
\\ \mathsf{L}_\mathcal{V}\psi^v=-\widetilde{\mathrm{div}}^v\,\mathbf{h}, \quad \mathsf{L}_\mathcal{V}\Psi^v=\widetilde{\mathrm{curl}}^v\,\mathbf{h}.
\end{gather}
\end{subequations}
The same vector can be represented over the dual network.  Setting $\mathbf{h}=\check{\mathbf{h}}\in\mathcal{L}$ (where $\check{}$ denotes representation on the dual network), the Helmholtz decomposition is given in terms of components parallel and perpendicular to links: thus $\check{\mathbf{h}}=\check{\mathbf{h}}^\parallel+\check{\mathbf{h}}^\perp$ where
\begin{align}
     \check{\mathbf{h}}^\parallel &= {\mathrm{grad}}^c \check{\psi}^c+\widetilde{\mathrm{CURL}}^v \check{\Psi}^v \in \mathcal{L}^\parallel, \\ 
     \check{\mathbf{h}}^\perp &=\widetilde{\mathrm{grad}}^v \check{\psi}^v+\mathrm{CURL}^c \check{\Psi}^c \in \mathcal{L}^\perp
    \label{eq:h2}
\end{align}
for some $\check{\psi}^c, \check{\Psi}^c \in \mathcal{C}$ and $\check{\psi}^v, \check{\Psi}^v\in\mathcal{T}$.   
The potentials are again determined from Poisson problems, namely
\begin{subequations}
\label{eq:divvcurlv}
\begin{gather}
\mathsf{L}_{\mathcal{T}}\check{\psi}^v=-{\mathrm{div}}^v\,\check{\mathbf{h}},
\quad
\mathsf{L}_{\mathcal{T}}\check{\Psi}^v=\mathrm{CURL}^v\,\check{\mathbf{h}}, \\
\mathsf{L}_\mathcal{C}\check{\psi}^c=-\widetilde{\mathrm{div}}^c\,\check{\mathbf{h}},\quad
\mathsf{L}_\mathcal{C}\check{\Psi}^c=\widetilde{\mathrm{CURL}}^c\,\check{\mathbf{h}}.
\end{gather}
\end{subequations}


In summary, (\ref{eq:h1}--\ref{eq:divvcurlv}) show how, given a vector field $\mathbf{h}$, we can determine the 8 corresponding scalar potentials that provide representations relative to the primal and dual networks (\ref{eq:h1}, \ref{eq:h2}), by solving a sequence of Poisson problems (\ref{eq:pot1}, \ref{eq:divvcurlv}) using the four Laplacians given in (\ref{eq:lapmat}).

\subsection{Potential theory for monolayers}
\label{sec:potthy}

\begin{figure*}
   \centering
    \includegraphics[width=\textwidth]{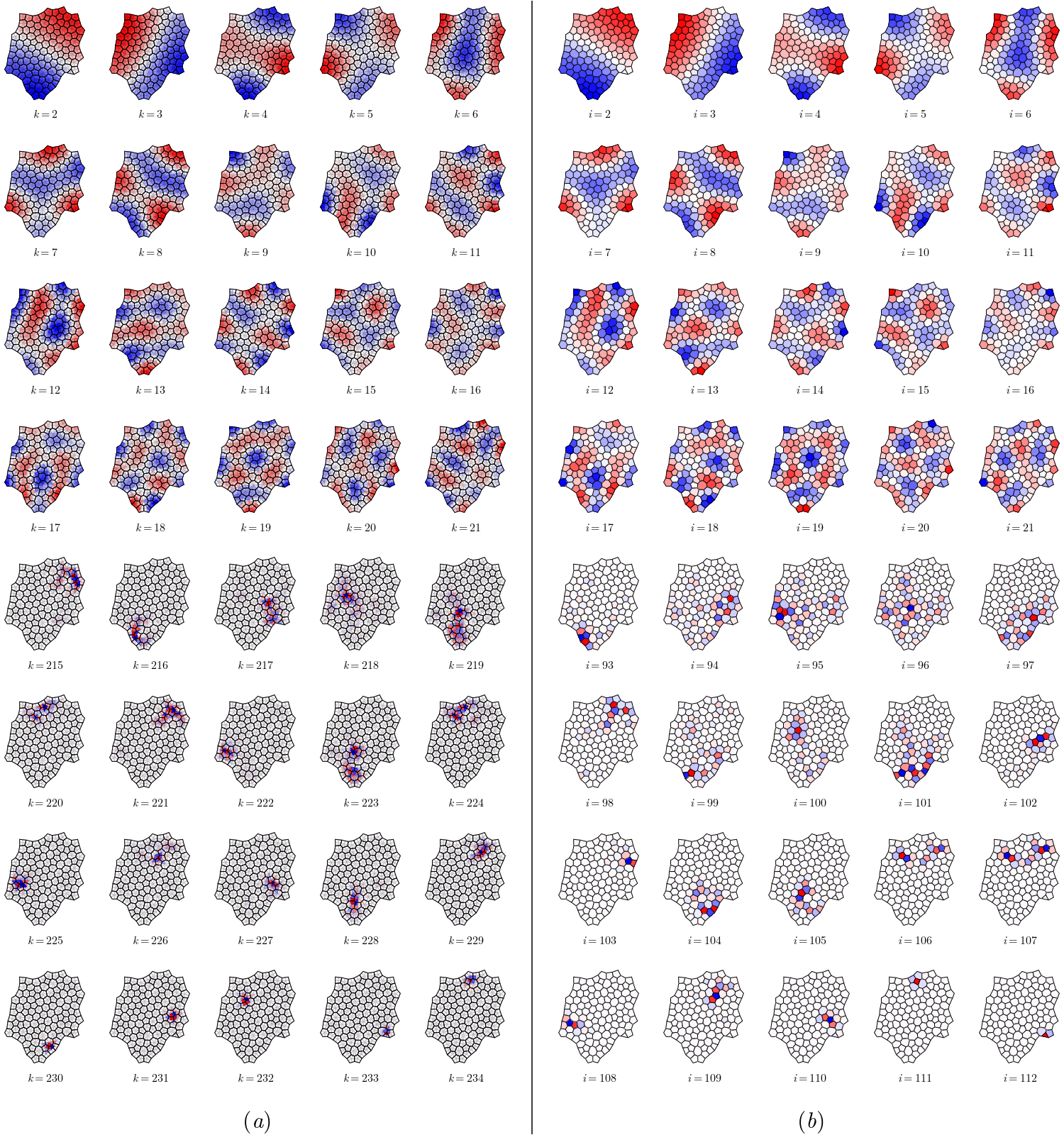}
    \caption{Eigenmodes of scalar Laplacians (a) $\mathsf{L}_\mathcal{V}$, defined over vertices, and (b) $\mathsf{L}_{\mathcal{F}}$, defined over cells, for the monolayer shown in Figure~\ref{fig:geometry}(a).  The first 20 spatially varying modes ($k=2,\dots,21$ and $i=2,\dots,21$, top) show clear resemblances; the final 20 modes $k=215,\dots,234$ and $i=93,\dots,112$ are distinct but are localised around common defects in the monolayer.}
    \label{fig:laps}
\end{figure*}

As demonstrated in Appendix~\ref{sec:invlap}, the four Laplacians are self-adjoint under suitable inner products.  The Poisson problems (\ref{eq:pot1}, \ref{eq:divvcurlv}) can then be evaluated directly using eigenmode decomposition, as demonstrated in (\ref{eq:lapsol}).  We illustrate the basis in which solutions can be expressed by plotting in Figure~\ref{fig:laps} the eigenmodes $e^{\mathcal{V}}_k$ of $\mathsf{L}_\mathcal{V}$ ($k=1,\dots,N_v$) and $e^{\mathcal{F}}_i$  of $\mathsf{L}_\mathcal{F}$ ($i=1,\dots,N_c$) for the monolayer shown in Figure~\ref{fig:geometry}(a).
$\mathsf{L}_{\mathcal{V}}$ has identical structure (but slightly different entries) to $\mathsf{L}_{\mathcal{T}}$ (likewise $\mathsf{L}_{\mathcal{C}}$ and $\mathsf{L}_{\mathcal{F}}$) and since links and edges are almost (but not exactly) orthogonal, the spectra and modes are qualitatively very similar.  Each Laplacian has a zero eigenvalue with a uniform eigenmode $\mathbbm{1}^c$ or $\mathbbm{1}^v$, where $\mathbbm{1}^v\equiv(1,1,\dots,1)$ has $N_v$ elements.  
The eigenmodes  of $\mathsf{L}_{\mathcal{V}}$ and $\mathsf{L}_{\mathcal{T}}$ are orthogonal under the inner product $[\cdot , \cdot]_{\mathcal{V}}$ and the eigenmodes of $\mathsf{L}_{\mathcal{F}}$ and $\mathsf{L}_{\mathcal{C}}$ are orthogonal under the inner product $[\cdot , \cdot]_{\mathcal{F}}$ (see (\ref{eq:inprod})).  Figure~\ref{fig:laps} (top) shows how the lower-order eigenmodes are influenced strongly by the shape of the monolayer but show consistent patterns over cells and vertices; the eigenvalues of the first 20 modes of $\mathsf{L}_\mathcal{V}$ and $\mathsf{L}_\mathcal{F}$ differ by no more than 8.2\%.  The highest-order modes exhibit strong localisation around defects in the monolayer (Figure~\ref{fig:laps}, bottom).

\section{Discrete forces, stresses and potentials}
\label{sec:3}

We now deploy this methodology to formulate discrete potentials of stress fields in a generic vertex model, specifically a vector force potential (Sec.~\ref{sec:forcepot}) and its underlying scalar potentials (Sec.~\ref{sec:stresspot}).  This reveals a couple stress, even when cells experience zero torque (Sec.~\ref{sec:pots}).

\subsection{Vector force potential}
\label{sec:forcepot}

\begin{figure}
    \centering
    \includegraphics[width=\columnwidth]{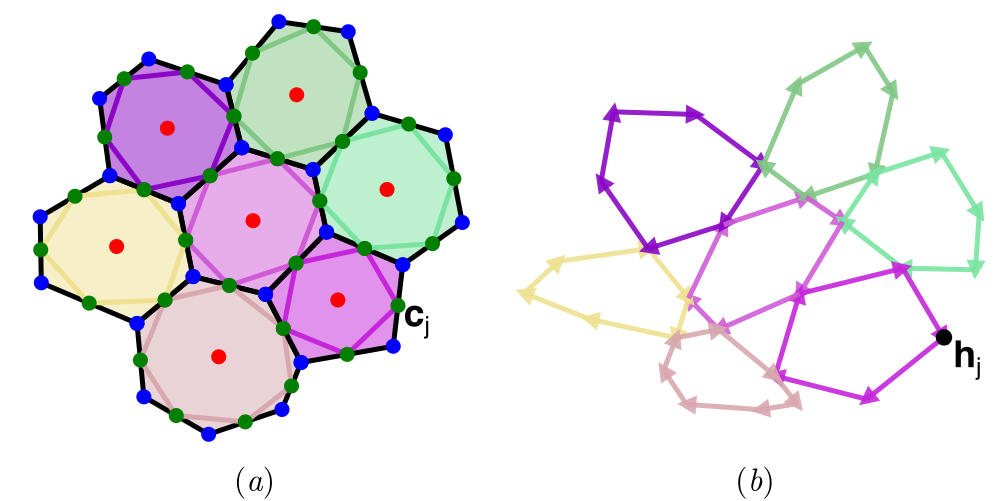}
    \caption{(a) The 7-cell cluster highlighted {\color{black}with bolder colours} in Figure~\ref{fig:geometry}(a), with lines linking edge centroids $\mathbf{c}_j$ (green dots); (b) the rotated force network of the highlighted cells, with arrow colours matching the cell applying the corresponding force. The vertices of the force network are force potentials $\mathbf{h}_j$. The topology and colours of the force network in (b) match those of the edge centroid network in (a). The data are from simulations (see Sec.~\ref{sec:comp}) with parameter values $\Gamma=0.2$, $L_0=0.75$, $P_{\mathrm{ext}}=0.2$.}
    \label{fig:forcenet}
\end{figure}

A standard computational implementation of the vertex model yields forces $\mathbf{f}_{ik}$ (of cell $i$, acting on vertex $k$) that balance at each vertex and around each cell, so that (respectively)

\begin{equation}
    \textstyle{\sum_i} C_{ik}\mathbf{f}_{ik}=\mathbf{0},\quad \textstyle{\sum_k} C_{ik}\mathbf{f}_{ik}=\mathbf{0}.
    \label{eq:for}
\end{equation}
These balances can be interpreted geometrically by rotating each force by $\pi/2$ (a form of Maxwell--Cremona force tiling \cite{bi2015a}), so that forces form closed triangles around each vertex (\ref{eq:for}a) and closed loops around each cell (\ref{eq:for}b), as illustrated in Figure~\ref{fig:forcenet}(b).  The network of rotated forces is topologically equivalent to the network of links connecting adjacent edge centroids \cite{jensen2020}, illustrated in Figure~\ref{fig:forcenet}(a).  Just as edge centroids $\mathbf{c}_j$ provide a vector potential for these links, so the vertices of the network of rotated forces define a vector potential $\mathbf{h}_j$ for rotated forces, via
\begin{equation}
    \mathbf{f}_{ik}=-{\textstyle \sum_j} \boldsymbol{\epsilon}_i B_{ij} \mathbf{h}_j A_{jk},
    \label{eq:rotf}
\end{equation}
such that $\mathbf{h}_j-\mathbf{h}_{j'}=\sum \boldsymbol{\epsilon}_i \mathbf{f}_{ik}$, summing over a path connecting vertex $j'$ to $j$ \cite{jensen2020}.  The force stress $\boldsymbol{\sigma}^c_i$ of cell $i$ (and the force stress $\boldsymbol{\sigma}_k^v$ defined over triangle $k$) can then be written as the first spatial moment of the forces acting on the cell (or triangle) \cite{ANB2018a}, or equivalently in terms of the force potential $\mathbf{h}$ \cite{jensen2020}, as
\begin{subequations}
\label{eq:stresses}
\begin{align}
    A_i \boldsymbol{\sigma}_i^c=&{\textstyle \sum_k} C_{ik}\mathbf{r}_k \otimes \mathbf{f}_{ik}=
    {\textstyle \sum_j} B_{ij} (\mathbf{t}_{j} \otimes \mathbf{h}_j) \boldsymbol{\epsilon}_i, \\
    E_k \boldsymbol{\sigma}_k^v=&
    -{\textstyle \sum_i} C_{ik}\mathbf{R}_i \otimes \mathbf{f}_{ik}={\textstyle{\sum_j}} A_{jk} (\mathbf{T}_j \otimes \mathbf{h}_j) \boldsymbol{\epsilon}_k.
\end{align}
\end{subequations}
Here differencing operators $\mathsf{A}$ and $\mathsf{B}$ in (\ref{eq:rotf})  have been transferred from $\mathbf{h}$ to $\mathbf{r}$ or $\mathbf{R}$ respectively, to express stresses in terms of edges and links.


The force-moment tensors in (\ref{eq:stresses}) are extensive and satisfy an important conservation principle  \cite{bi2015a}: summing $A_i \boldsymbol{\sigma}^c_i$ over adjacent cells yields a quantity defined entirely by forces or force potentials at the periphery of the cells (because of cancellation of internal forces for a system in equilibrium).  In particular,
\begin{equation}
    {\textstyle \sum_i} A_i {\boldsymbol{\sigma}}_i^c=-P_{\mathrm{ext}}\mathsf{I} \mathcal{A}
    \label{eq:outerbc}
\end{equation}
for a monolayer of total area $\mathcal{A}$ that is subject to a uniform external pressure $P_{\mathrm{ext}}$ \cite{ANB2018a}.  The periphery of the force network is defined by rotated peripheral forces: as these are assumed to act normal to peripheral edges, the rotated peripheral forces form a closed loop having exactly the same shape as the periphery of the monolayer, but scaled by $P_\mathrm{ext}$ \cite{jensen2020}.

The conditions that force-stresses have zero divergence can be demonstrated by splitting stresses over cells and triangles into elements associated with edges and links, as explained in Appendix~\ref{sec:micro}.  Dividing stresses into isotropic and antisymmetric components (Appendix~\ref{sec:micro}) also reveals that
\begin{subequations}
\label{eq:pef}
\begin{align}
P_{\mathrm{eff},i}\equiv &\tfrac{1}{2}  \mathrm{Tr}(\boldsymbol{\sigma}_i^c)=-\tfrac{1}{2} \{ \mathrm{div}^c\, \mathbf{h} \}_i
,\\
P_{\mathrm{eff},k}\equiv &\tfrac{1}{2} \mathrm{Tr}(\boldsymbol{\sigma}_k^v)=-\tfrac{1}{2} \{ \mathrm{div}^v\, \mathbf{h} \}_k
,\\
 \boldsymbol{\sigma}_i^{c(a)}=&\tfrac{1}{2}\boldsymbol{\epsilon}_i \{ \mathrm{curl}^c\, {\mathbf{h}} \}_i,
\label{eq:peffi} \\
\boldsymbol{\sigma}_k^{v(a)}=&\tfrac{1}{2}\boldsymbol{\epsilon}_k \{  \mathrm{CURL}^v\, {\mathbf{h}}\}_k.
\label{eq:peffk}
\end{align}
\end{subequations}
Thus the projections of the vector force potential $\mathbf{h}$ onto edges and links (in curls, (\ref{eq:pef}c,d)) are associated with couples on cells and triangles, whereas the projections onto normals to cells and triangles (in divergences, (\ref{eq:pef}a,b)) contribute to the isotropic stresses, which we express as the effective pressure $P_\mathrm{eff}$.

\subsection{Scalar stress potentials}
\label{sec:stresspot}

We now pursue the discrete analogue of (\ref{eq:curlcurl}), expressing the force stress in terms of scalar potentials and identifying an associated couple-stress vector.   The force potential $\mathbf{h}$ can be expressed in terms of scalar potentials using (\ref{eq:h1}) and (\ref{eq:h2}), so that (\ref{eq:pef}) becomes
\begin{subequations}
\label{eq:tra}
\begin{align}
 \mathrm{Tr}(\boldsymbol{\sigma}_i^c)& = \{ \mathsf{L}_\mathcal{F} \psi^c \}_i, &
 \boldsymbol{\sigma}_i^{c(a)}&=\tfrac{1}{2} \boldsymbol{\epsilon}_i  \{ \mathsf{L}_\mathcal{F} \Psi^c \}_i, \\
 \mathrm{Tr}(\boldsymbol{\sigma}_k^v)&= \{ \mathsf{L}_\mathcal{T} \check{\psi}^v \}_k , &
\boldsymbol{\sigma}_k^{v(a)}&=\tfrac{1}{2}\boldsymbol{\epsilon}_k \{ \mathsf{L}_\mathcal{T} \check{\Psi}^v \}_k.
\end{align}
\end{subequations}
We construct vectors orthogonal to $\mathbf{h}$ and $\check{\mathbf{h}}$  (its dual representation) such that
\begin{subequations}
\label{eq:hpot}
\begin{align}
-\boldsymbol{\epsilon}_i \mathbf{h}=&(\mathrm{curl}^v \psi^v - \widetilde{\mathrm{grad}}^c \Psi^c) \nonumber\\
& \qquad + (\widetilde{\mathrm{curl}}^c \psi^c -\mathrm{grad}^v \Psi^v),\\
-\boldsymbol{\epsilon}_k \check{\mathbf{h}}=&(\mathrm{CURL}^c \check{\psi}^c - \widetilde{\mathrm{grad}}^v \check{\Psi}^v)\nonumber \\ &\qquad +(\widetilde{\mathrm{CURL}}^v \check{\psi}^v-\mathrm{grad}^c\check{\Psi}^c).
\end{align}
\end{subequations}
Then, by analogy with (\ref{eq:sigmah}), we can re-write (\ref{eq:stresses}) as
\begin{equation}
\boldsymbol{\sigma}^c=\mathrm{curl}^c \otimes (-\boldsymbol{\epsilon}_i\mathbf{h}),\quad
\boldsymbol{\sigma}^v=\mathrm{CURL}^v \otimes (-\boldsymbol{\epsilon}_k\check{\mathbf{h}}).
\label{eq:dfp}
\end{equation}
This representation is verified in Appendix~\ref{sec:sdp}.  Eq.~(\ref{eq:dfp}) shows how the Cauchy stresses are defined in terms of the force potential $\mathbf{h}$, which is given in turn in terms of eight potentials (four per network) in (\ref{eq:hpot}).  The relationship with (\ref{eq:curlcurl}a) becomes clear:
\begin{subequations}
\begin{align}
\boldsymbol{\sigma}^c=&\mathrm{curl}^c \otimes \big[ (\mathrm{curl}^v \psi^v - \widetilde{\mathrm{grad}}^c \Psi^c)_\perp  \nonumber \\ &\qquad + (\widetilde{\mathrm{curl}}^c \psi^c -\mathrm{grad}^v \Psi^v)_\parallel \big], \\
\boldsymbol{\sigma}^v=&\mathrm{CURL}^v \otimes \big[
(\mathrm{CURL}^c \check{\psi}^c - \widetilde{\mathrm{grad}}^v \check{\Psi}^v)_\perp \nonumber \\
&\qquad +(\widetilde{\mathrm{CURL}}^v \check{\psi}^v-\mathrm{grad}^c\check{\Psi}^c)_\parallel
\big],
\end{align}
\end{subequations}
where subscripts $\perp$ and $\parallel$ serve as reminders of the orientations of the vectors relative to edges (for $\boldsymbol{\sigma}^c$) and links (for $\boldsymbol{\sigma}^v$) respectively.  We note also from (\ref{eq:hpot}) that
$\mathbf{h}=\mathrm{grad}^v \psi^v + \widetilde{\mathrm{grad}}^c \psi^c -\boldsymbol{\mu}$ or $ \check{\mathbf{h}}=\mathrm{grad}^c \check{\psi}^c  +\widetilde{\mathrm{grad}}^v \check{\psi}^v-\check{\boldsymbol{\mu}}$, where
\begin{subequations}
\label{eq:hmu}
\begin{align}
\boldsymbol{\mu} =& - \widetilde{\mathrm{curl}}^c \Psi^c - \mathrm{curl}^v \Psi^v,\\
\check{\boldsymbol{\mu}} =& - \widetilde{\mathrm{CURL}}^v \check{\Psi}^v -\mathrm{CURL}^c\check{\Psi}^c,
\end{align}
\end{subequations}
suggesting how ($\Psi^v$, $\Psi^v$) and $(\check{\Psi}^v, \check{\Psi}^c)$ can be interpreted as scalar potentials for the the candidate couple-stress vector $\boldsymbol{\mu}$, in its representation over cells and over triangles respectively (the analogue of (\ref{eq:curlcurl}b)).  We note that $\mathrm{curl}^c \boldsymbol{\mu} =-\mathsf{L}_{\mathcal{F}} \Psi^c$ and $\mathrm{CURL}^v \check{\boldsymbol{\mu}} =-\mathsf{L}_{\mathcal{T}} \check{\Psi}^v$ so that
\begin{equation}
\boldsymbol{\sigma}^{c(a)}=-\tfrac{1}{2}{\boldsymbol{\epsilon}}_i \mathrm{curl}^c \boldsymbol{\mu}, \quad
\boldsymbol{\sigma}^{v(a)}=-\tfrac{1}{2}{\boldsymbol{\epsilon}}_k \mathrm{CURL}^v \boldsymbol{\mu}.
\label{eq:antistress}
\end{equation}

\subsection{Potentials with zero couple on cells}
\label{sec:pots}

We now specialise to the case when individual cells experience no couple (the case relevant to the standard vertex model). 
To do so, we require $\mathsf{L}_\mathcal{F} \Psi^c=0$ in (\ref{eq:tra}a), which implies $\Psi^c =a\mathbbm{1}^c$ for some constant $a$, which we can take to be zero without loss of generality.  Thus $\Psi^c=0$. Then $\mathrm{curl}^c\, \mathbf{h}$ vanishes (so that $\boldsymbol{\sigma}_i^{(a)}=\mathsf{0}$ in (\ref{eq:pef})) but $\mathrm{CURL}^v \,\mathbf{h} = \mathsf{L}_{\mathcal{T}} \check{\Psi}^v$ is likely to be non-zero (giving non-zero torque on triangles).  We identify (from (\ref{eq:hmu}a)) the couple stress vector $\boldsymbol{\mu}$ with $-\mathrm{curl}^v \Psi^v$, which is normal to edges, satisfying $\mathrm{div}^c \boldsymbol{\mu}=\mathbf{0}$.  In general we expect $\boldsymbol{\mu}\equiv \check{\boldsymbol{\mu}}$ to be described by non-zero $\check{\Psi}^v$ and $\check{\Psi}^c$ in (\ref{eq:hmu}b).

To ensure that the Poisson problems such as (\ref{eq:pot1}a, \ref{eq:divvcurlv}a) have solutions, we integrate the forcing terms over the monolayer.
Making use of (\ref{eq:outerbc}), (\ref{eq:pef}) and (\ref{eq:stokes}), we find
\begin{subequations}
\label{eq:spav}
\begin{align}
    [\mathbbm{1}^c,-\mathrm{div}^c\,\mathbf{h}]_\mathcal{F}=&2[\mathbbm{1}^c,P_{\mathrm{eff} i}]_\mathcal{F} \nonumber \\\equiv & 2 {\textstyle{\sum_i}} A_i P_{\mathrm{eff}i} = 2 \mathcal{A} P_{\mathrm{ext}}, \\
    [\mathbbm{1}^v,-\mathrm{div}^v\,\mathbf{h}]_\mathcal{V}=&2[\mathbbm{1}^v,P_{\mathrm{eff} i}]_\mathcal{V}\nonumber \\ \equiv & 2 {\textstyle{\sum_i}} E_k P_{\mathrm{eff}k} = 2 \mathcal{A} P_{\mathrm{ext}},\\
    [\mathbbm{1}^c,\mathrm{curl}^c\,\mathbf{h}]_\mathcal{F}=&\, 0, \\
    [\mathbbm{1}^v,\mathrm{CURL}^v\,\mathbf{h}]_\mathcal{V}=&-[\mathbbm{1}^v,\mathrm{CURL}^v \,\boldsymbol{\mu} ]_{\mathcal{V}} =0.
    \end{align}
\end{subequations}
Writing $\mathbf{h}=-P_{\mathrm{ext}}\mathbf{m}+\breve{\mathbf{h}}$, where $\mathbf{m}_j$ is the intersection of edge $\mathbf{t}_j$ with link $\mathbf{T}_j$, the identities $\tfrac{1}{2} \mathrm{div}^c\,\mathbf{m}=\mathbbm{1}^c$, $\tfrac{1}{2} \mathrm{div}^v\,\mathbf{m}=\mathbbm{1}^v$ (from (\ref{eq:m})) imply that $\mathrm{div}^c\,\mathbf{h}=-2P_{\mathrm{ext}}\mathbbm{1}^c+\mathrm{div}^c \,\breve{\mathbf{h}}$ with $[\mathbbm{1}^c,\mathrm{div}^c\,\breve{\mathbf{h}}]_\mathcal{F}=0$ and $\mathrm{div}^v\,\mathbf{h}=-2P_{\mathrm{ext}}\mathbbm{1}^v+\mathrm{div}^v \,\breve{\mathbf{h}}$ with $[\mathbbm{1}^v,\mathrm{div}^v\,\breve{\mathbf{h}}]_\mathcal{V}=0$.  Thus $\breve{\mathbf{h}}$ satsifies the solvability conditions necessary to invert $\mathcal{L}_\mathcal{F}$ for $\psi^c$ and $\mathcal{L}_\mathcal{T}$ for $\check{\psi}^v$.   
Thus, on the primal network, we solve $\mathsf{L}_\mathcal{F} \psi^c=-\mathrm{div}^c\, \breve{\mathbf{h}} =-\mathrm{div}^c\, (\mathbf{h}+P_\mathrm{ext} \mathbf{m})$ to determine $\psi^c$.
Similarly, on the dual network, we solve $\mathsf{L}_\mathcal{T} \check{\psi}^v=-\mathrm{div}^v \,\breve{\mathbf{h}}=-\mathrm{div}^v\,(\mathbf{h}+P_\mathrm{ext}\mathbf{m})$ to determine $\check{\psi}^v$.
When edges and links are almost orthogonal, we find that $\mathsf{L}_\mathcal{C}\approx \mathsf{L}_\mathcal{F}$ and $\mathsf{L}_\mathcal{V}\approx \mathsf{L}_\mathcal{T}$ (in that the spectra and eigenmodes are almost identical) 
so that 
$\check{\psi}^c\approx \psi^c$, $\check{\psi}^v\approx \psi^v$ and  $\check{\Psi}^v\approx -\Psi^v$.  We therefore illustrate just three of the seven non-zero potentials below. 

\section{Energetics}
\label{sec:const}

We now introduce a constitutive model by specifying the general form of the mechanical energy, introducing the principle of virtual work for pure strain (Sec.~\ref{sec:vw}) and for {\color{black}in-plane} bending deformations (Sec.~\ref{sec:bend}).

\subsection{Virtual work in the vertex model}
\label{sec:vw}

We introduce an energy per cell $U_i=U(A_i,L_i)$ (assuming cells have homogeneous mechanical properties) and define a pressure and tension as $\mathcal{P}_i\equiv \partial U/\partial A_i$ and $\mathcal{T}_i\equiv \partial U/\partial L_i$ respectively.  The total energy of the monolayer is $\mathcal{U}=\sum_i U_i + P_{\mathrm{ext}} \mathcal{A}$, where $\mathcal{A}=\sum_i A_i$ and $P_{\mathrm{ext}}$ is an external pressure applied to the periphery of the monolayer, as in (\ref{eq:outerbc}).  We therefore assume there is no moment traction at the monolayer periphery and only a normal force traction.  $\mathcal{U}$ is a function of vertex locations, via the dependence of areas and perimeters on $\mathbf{r}_k$.  Suppose the monolayer is in a stationary equilibrium configuration (denoted with a prime) and consider virtual displacements $\delta \mathbf{r}_k$ of its vertices.  The expansion
\begin{multline}
\mathcal{U}=\mathcal{U}'+{\sum_{i,k}}\left( \frac{\partial U_i}{\partial \mathbf{r}_k} +P_{\mathrm{ext}} \frac{\partial A_i}{\partial \mathbf{r}_k}  \right)' \cdot \delta \mathbf{r}_k  \\ +{\sum_{i,k,k^\star}} \tfrac{1}{2} \delta \mathbf{r}_k \cdot \left(\frac{\partial^2 U_i}{\partial \mathbf{r}_k \partial \mathbf{r}_{k^\star}} + P_{\mathrm{ext}} \frac{\partial^2 A_i}{\partial \mathbf{r}_k \partial \mathbf{r}_{k^\star}}  \right)' \cdot \delta \mathbf{r}_{k^\star}+\dots
\label{eq:ux}
\end{multline}
reveals the force $\partial U_i/\partial \mathbf{r}_k$ exerted at vertex $k$ by cell $i$.  Given that $\partial A_i/\partial \mathbf{r}_k =\tfrac{1}{2} \mathbf{n}_{ij} \overline{A}_{jk}$, it follows that $\sum_i \partial A_i/\partial \mathbf{r}_k$ vanishes at all internal tricellular junctions, so that $P_{\mathrm{ext}}$ contributes to forces only along the monolayer's periphery, via virtual displacement of edge centroids:
\begin{align*}
P_{\mathrm{ext}} {\textstyle \sum_{i,k}} \left( \frac{\partial A_i}{\partial \mathbf{r}_k} \right)' \cdot \delta \mathbf{r}_k
=& \tfrac{1}{2}P_{\mathrm{ext}} {\textstyle \sum_{i,j,k} } \overline{A}_{jk} \mathbf{n}_{ij}'\cdot \delta\mathbf{r}_k \nonumber \\
=&  P_{\mathrm{ext}} {\textstyle \sum_{i,j} }  \mathbf{n}_{ij}'\cdot \delta\mathbf{c}^p_j.
\end{align*}
The principle of virtual work requires that, in equilibrium, $\mathcal{U}$ is unchanged by small  independent displacements of each of the vertices. Equivalently, the sum of all forces at each vertex vanishes when the monolayer is at an equilibrium, \hbox{i.e.} $\sum_{i} C_{ik} (\partial U_i/\partial \mathbf{r}_k+P_{\mathrm{ext}} \partial A_i/\partial \mathbf{r}_k)'=\mathbf{0}$ for all $k$, as in (\ref{eq:for}).  The second variation in (\ref{eq:ux}) captures weakly nonlinear effects and establishes the stability or otherwise of the equilibrium \cite{yan2019}, including any jamming/unjamming transition \cite{bi2015}.  We work below with the first variation, but consider how the forces organise into stresses acting over cells.

Consider variations that can be expressed as a smooth function of position under a deformation $\mathbf{u}(\mathbf{x})$, \hbox{i.e.} we map vertices from $\mathbf{r}_k'$ to $\mathbf{r}_k=\mathbf{r}_k'+\mathbf{u}(\mathbf{r}_k')$  so that $\delta \mathbf{r}_k=\mathbf{u}(\mathbf{r}_k')$.  Suppose first that $\mathbf{u}$ is linear in $\mathbf{x}$, so that the virtual displacements are of the form $\delta\mathbf{r}_k=\mathbf{u}_0+(\mathsf{E}+\mathsf{W})\cdot (\mathbf{r}_k'-\mathbf{x}_0)$, where $\mathsf{E}=\mathsf{E}^\top$ and $\mathsf{W}=-\mathsf{W}^\top$ are a small uniform strain and rotation respectively and $\mathbf{u}_0$ and $\mathbf{x}_0$ are constants.  The principle of virtual work can be formulated by noting that variations in cell area and perimeter under this deformation are \cite{ANB2018b}
\begin{equation}
    A_i=A_i'[1+\mathsf{I}:\mathsf{E}+\dots], \quad L_i=L_i'[1+\mathsf{Q}_i:\mathsf{E}+\dots]
    \label{eq:allin}
\end{equation}
where $L_i' \mathsf{Q}_i\equiv\sum_j \overline{B}_{ij} \mathbf{t}_j'\otimes \hat{\mathbf{t}}_j'$, with no contribution from uniform translation ($\mathbf{u}_0$, $\mathbf{x}_0$) and rotation ($\mathsf{W}$).  Then the first variation in energy can be written
\begin{multline*}
    \mathcal{U}=\mathcal{U}'+{\textstyle \sum_i} \left[\mathcal{P}_i (A_i-A_i')+\mathcal{T}_i (L_i-L_i')+\dots\right] \\
    + P_\mathrm{ext}(\mathcal{A}-\mathcal{A}').
\end{multline*}
Thus $ \mathcal{U}=\mathcal{U}'+{\textstyle \sum_i} A_i' \boldsymbol{\varsigma}^{c} : \mathsf{E} + P_\mathrm{ext} \mathcal{A}' \mathsf{I}:\mathsf{E}+\dots,$ where
\begin{equation}
\boldsymbol{\varsigma}_i^{c}\equiv \mathcal{P}_i'\mathsf{I}+\frac{\mathcal{T}_i' L_i'}{A_i'} \mathsf{Q}_i.
\label{eq:varstress}
\end{equation}
This reveals the leading-order cell force-stress tensor $\boldsymbol{\varsigma}_i^{c}$ that is energy-conjugate to $\mathsf{E}$ and symmetric.  The virtual work principle (\hbox{i.e.}, at equilibrium, the energy does not vary for small but arbitrary strains $\mathsf{E}$) recovers the bulk constraint $(\ref{eq:outerbc})$ with $\boldsymbol{\sigma}^c=\boldsymbol{\varsigma}^c$. The isotropic component of $\boldsymbol{\varsigma}^c$ gives the cell effective pressure as \cite{ANB2018a}
\begin{equation}
    P_{\mathrm{eff},i}=\mathcal{P}_i+\frac{\mathcal{T}_i L_i}{2A_i}.
    \label{eq:peff}
\end{equation}
Comparison to direct evaluation of $\boldsymbol{\sigma}^c$ as a first moment of forces in (\ref{eq:stresses}) \cite{ANB2018a} shows the success of the affine approximation in this instance.

\subsection{Strain gradients}
\label{sec:bend}

We now take this argument a step further and consider displacements involving gradients of strain, allowing $\mathbf{u}$ to be quadratic in $\mathbf{x}$.  We continue to  neglect effects that are quadratic in strain but account for first- and second-order deformation gradients $\nabla \mathbf{u}$ and $\mathsf{M}\equiv (\nabla\otimes \nabla) \mathbf{u}$,
and reformulate the first variation in (\ref{eq:ux}) in terms of $\mathsf{E}\equiv \tfrac{1}{2}(\nabla \mathbf{u}+\nabla \mathbf{u}^\top)$, $\nabla\mathsf{E}$ and $\boldsymbol{\kappa}\equiv -\tfrac{1}{4}(\nabla^2 \mathbf{u} -\nabla(\nabla \cdot\mathbf{u}))= -\tfrac{1}{2} \mathsf{I}:\nabla\mathsf{W}$.  We interpolate deformation gradients evaluated on vertices onto edge centroids and cell centres, using Taylor expansion to capture the leading-order effect of spatial variations across any single cell.  Accordingly, we use subscripts $i$, $j$ and $k$ to describe fields evaluated at cell centres, edge centroids and vertices, writing $\mathbf{u}_i\equiv \mathbf{u}(\mathbf{R}_i')$, $\mathbf{u}_j\equiv \mathbf{u}(\mathbf{c}_j')$ and $\mathbf{u}_k=\mathbf{u}(\mathbf{r}_k')$ and so on.  We retain second derivatives of $\mathbf{u}$ but discard third and higher derivatives, assuming deformations vary over scales long compared to the size of individual cells.  As shown in Appendix~\ref{sec:ap}, the changes in cell perimeter and area to this order become
\begin{subequations}
\label{eq:la}
\begin{align}
L_i=& L_i'\left[1+\mathsf{Q}_i : \mathsf{E}_i+
\mathsf{X}_i  \mathrel{\lower.03em\hbox{\rlap{$\cdot$}}{:}} (\mathsf{\nabla E})_i\right], \\
A_i=&A_i'\Big[ 1+\mathsf{I}:\mathsf{E}_i + \tfrac{1}{2} (\boldsymbol{\rho}_i'-\mathbf{R}_i')\cdot \nabla (\mathsf{I}:\mathsf{E})_i  \nonumber \\
& \qquad + \mathsf{Y}_i \mathrel{\lower.03em\hbox{\rlap{$\cdot$}}{:}} (\nabla \mathsf{E})_i - \left[ \frac{1}{6A_i'}\textstyle{\sum_j} (t_j')^2 \mathbf{n}_{ij}'\right] \cdot \boldsymbol{\kappa}_i \Big].
\end{align}
\end{subequations}
In comparison to (\ref{eq:allin}), we note additional terms.  That involving $\boldsymbol{\rho}_i'-\mathbf{R}_i'$ gives a correction pushing the evaluation of $\mathsf{E}_i$ towards the cell area centroid $\boldsymbol{\rho}_i$.  The third-order tensors $\mathsf{X}_i$ and $\mathsf{Y}_i$ (see (\ref{eq:xi}, \ref{eq:yi})) characterise the impact of strain gradients on cell perimeter and area respectively.  They are size-dependent, as is appropriate for objects that measure a gradient.  $\nabla \mathsf{W}$ does not change perimeter to this order, but it alters cell area through the curvature $\boldsymbol{\kappa}_i$.
Returning to (\ref{eq:ux}), the energy maps from $\mathcal{U}_0\equiv \mathcal{U}'+P_{\mathrm{ext}} \mathcal{A}'$  to
 \begin{align}
&\mathcal{U}_0+{\textstyle{\sum_i}}[\mathcal{P}_i'  (A_i-A_i')+\mathcal{T}_i' (L_i-L_i')]\nonumber \\
& \qquad -P_{\mathrm{ext}}{\textstyle \sum_{i,j}}\mathbf{n}_{ij}'\cdot (\mathbf{c}^p_j-(\mathbf{c}^{p}_j)') +\dots\nonumber \\
&=\mathcal{U}_0+\textstyle{\sum_i}\Big\{ A_i' (\boldsymbol{\varsigma}_i^{c}  +P_{\mathrm{ext}}\mathsf{I}):\mathsf{E}_i \nonumber \\ & \qquad + \left[(\mathcal{P}_i'+P_\mathrm{ext}) A_i' \mathsf{Y}_i +\mathcal{T}_i' L_i' \mathsf{X}_i\right]  \mathrel{\lower.03em\hbox{\rlap{$\cdot$}}{:}}  \nabla\mathsf{E}_i \nonumber \\
& \qquad + \tfrac{1}{2}(\mathcal{P}_i' +P_\mathrm{ext}) A_i' (\boldsymbol{\rho}_i-\mathbf{R}_i)\cdot \nabla (\mathsf{I}:\mathsf{E})_i \nonumber \\
& \qquad -2A_i' \boldsymbol{\mathfrak{m}}_i^c \cdot\boldsymbol{\kappa}_i \Big\}
 +\dots,
\label{eq:cellen}
 \end{align}
using (\ref{eq:la}) and neglecting quantities that are quadratic in strains, where 
 \begin{equation}
\boldsymbol{\mathfrak{m}}_i^c\equiv \frac{\mathcal{P}_i'+P_\mathrm{ext}}{12A_i'}\textstyle{\sum_j} (t_j')^2 \mathbf{n}_{ij}'.
\label{eq:cstress}
 \end{equation}
Now $\mathbf{n}_{ij} \cdot (\mathbf{c}_j^p - \mathbf{c}^{p'}_j) =\mathbf{n}_{ij} \cdot \mathbf{u}^p_j$ to this order (by (\ref{eq:cj})), so that $P_{\mathrm{ext}}$ does not exert any moment on the periphery. 
(For second-gradient materials, terms that are energy-conjugate to gradients of $\mathsf{E}$ and $\mathsf{W}$ are identified as hyperstresses \cite{toupin1964}).  However direct comparison of (\ref{eq:cellen}) with the principle of virtual work for a continuum (\ref{eq:vw}) is not straightforward: deformations for which $\nabla\mathsf{E}$ are eliminated but which retain $\boldsymbol{\kappa}$ are not possible via compatibility constraints, and the present affine (or Cauchy--Born) approximation does not account for possible local adjustments of vertex locations that ensure equilibration.  
Nevertheless, the comparison suggests $\boldsymbol{\mathfrak{m}}_i^c$ as a candidate couple-stress vector, defined over cells.
Given that $\sum_j \mathbf{n}_{ij}'=\mathbf{0}$, $\boldsymbol{\mathfrak{m}}^c_i$ vanishes for symmetric cells, for which $t_j'$ is uniform.   

We can repartition the contribution to the energy associated with $\boldsymbol{\mathfrak{m}}_i^c$ to define its analogue on edges and links.  As gradients in curvature across the monolayer will not play a role in what follows, we take $\boldsymbol{\kappa}$ to be uniform, and drop primes, to define the vector $\boldsymbol{\mathfrak{m}}_j$ attributed to edges via ${\textstyle \sum_i} A_i (2 \boldsymbol{\mathfrak{m}}_i^c \cdot \boldsymbol{\kappa})= {\textstyle \sum_j} \tfrac{1}{2} F_j (2\boldsymbol{\mathfrak{m}}_j \cdot \boldsymbol{\kappa})$ where
\begin{equation}
\boldsymbol{\mathfrak{m}}_j=\frac{t_j^2}{6 F_j} {\textstyle \sum_i} (\mathcal{P}_i+P_\mathrm{ext}) \mathbf{n}_{ij},
\label{eq:edgecouple1}
\end{equation}
where area is partitioned into trapezia of area $\tfrac{1}{2} F_j$, associated with edge/link $j$. $P_\mathrm{ext}$ makes zero contribution to $\boldsymbol{\mathfrak{m}}$ at all internal edges, but contributes along peripheral edges.  $\boldsymbol{\mathfrak{m}}$ has zero curl around cells (because it acts along normals to edges and sits in $\mathcal{E}^\perp$) but has non-zero curl around triangles of the dual network: for internal vertices,
\begin{multline}
\mathcal{C}_k\equiv \{\mathrm{CURL}^v\,\boldsymbol{\mathfrak{m}}\}_k=E_k^{-1}{\textstyle \sum_j} A_{jk} \mathbf{T}_j\cdot \boldsymbol{\mathfrak{m}}_j \\
= - \frac{1}{6 E_k} {\textstyle \sum_{i,j}} \mathcal{P}_i B_{ij} t_j^2 A_{jk}.
\label{eq:tricouple}
\end{multline}
Here, the pressure difference across edge $j$, $\sum_i B_{ij}\mathcal{P}_i$, is multiplied by $t_j^2$ to give a moment, and the three contributions to the moment at the tricellular junction are summed at the vertex.    $\mathcal{C}_k$ vanishes if pressures are uniform ($\sum_i \mathcal{P}_i B_{ij}=0$) or if the edges are of uniform size (because $\mathsf{B}\mathsf{A}=\mathsf{0}$).  We do not seek to impose any conditions on $\mathrm{div}^c\,\boldsymbol{\mathfrak{m}}$ or $\mathrm{div}^v\,\boldsymbol{\mathfrak{m}}$.

To summarise, we now have two representations of couple stress.  The vector $\boldsymbol{\mu}$ in (\ref{eq:hmu}) is associated with the contribution of the vector force potential associated with curls.  It is normal to cell edges (sitting in $\mathcal{E}^\perp$, so exerting zero couple traction on any cell because $\mathbf{n}_{ij}\cdot \boldsymbol{\varepsilon}\boldsymbol{\mu}_j=0$), contributes to torques around cell vertices via (\ref{eq:antistress}) and sums to zero over the monolayer via (\ref{eq:spav}d).  However it is not energy-conjugate to the curvature $\boldsymbol{\kappa}$, as shown in (\ref{eq:conj}).  In contrast, the vector $\boldsymbol{\mathfrak{m}}$ also sits in $\mathcal{E}^\perp$, is energy conjugate to $\boldsymbol{\kappa}$ via (\ref{eq:cellen}) but is not expressible as a curl of a scalar field defined on vertices. $\boldsymbol{\mathfrak{m}}$ has a direct physical interpretation as pressure differences acting over cell edges meeting at a vertex to generate a torque, but was derived under an affine approximation that needs evaluation.


\section{Computations}
\label{sec:comp}

We implemented the vertex model using the commonly-used cell energy
\begin{equation}
U(A_i,L_i)=\tfrac{1}{2} (A_i-1)^2 + \tfrac{1}{2} \Gamma (L_i -L_0)^2
\label{eq:energy}
\end{equation}
for which cell pressure and tension are linear in area and perimeter: $\mathcal{P}_i=A_i-1$ and $\mathcal{T}_i=\Gamma(L_i-L_0)$.  A vertex drag was implemented so that the system could relax to equilibrium under 
\begin{equation}
\eta \mathrm{d}\mathbf{r}_k/\mathrm{d}t=-{\textstyle\sum_i} C_{ik} \mathbf{f}_{ik}
\label{eq:dynamics}
\end{equation}
for some $\eta>0$.  We chose $\Gamma=0.2$ and $L_0=0.75$, values for which the monolayer is in a jammed state \cite{farhadifar2007, bi2015}.  An isolated monolayer under uniform external pressure $P_{\mathrm{ext}}$ was established by starting with a small number of cells and allowing cell divisions to occur randomly for an interval; {\color{black} we examine configurations in which the monolayer has settled}  to an equilibrium state (Figure~\ref{fig:geometry}a, Appendix~\ref{sec:compimp}).

The forces $\mathbf{f}_{ik}$ acting at each vertex in the equilibrium state were rotated and assembled to form a force network, as illustrated in Figure~\ref{fig:forcenet}(b) for the cluster of cells shown in Figure~\ref{fig:forcenet}(a).  The three rotated forces around each internal vertex form a closed triangle, and the $Z_i$ forces around cell $i$ form closed loops, confirming (\ref{eq:for}).  
For sufficiently large $\vert P_{\mathrm{ext}}\vert $ the force network may form a planar graph.  However in general this is not the case, although the force network maintains the same topology as that of connections between adjacent edge centroids (Figure~\ref{fig:forcenet}a) \cite{jensen2020}.  The distorted force loops provide a striking illustration of spatially and temporally heterogeneous loading experienced by individual cells as the monolayer grows (Movie 1).

\begin{figure}
    \centering
    \includegraphics[width=\columnwidth]{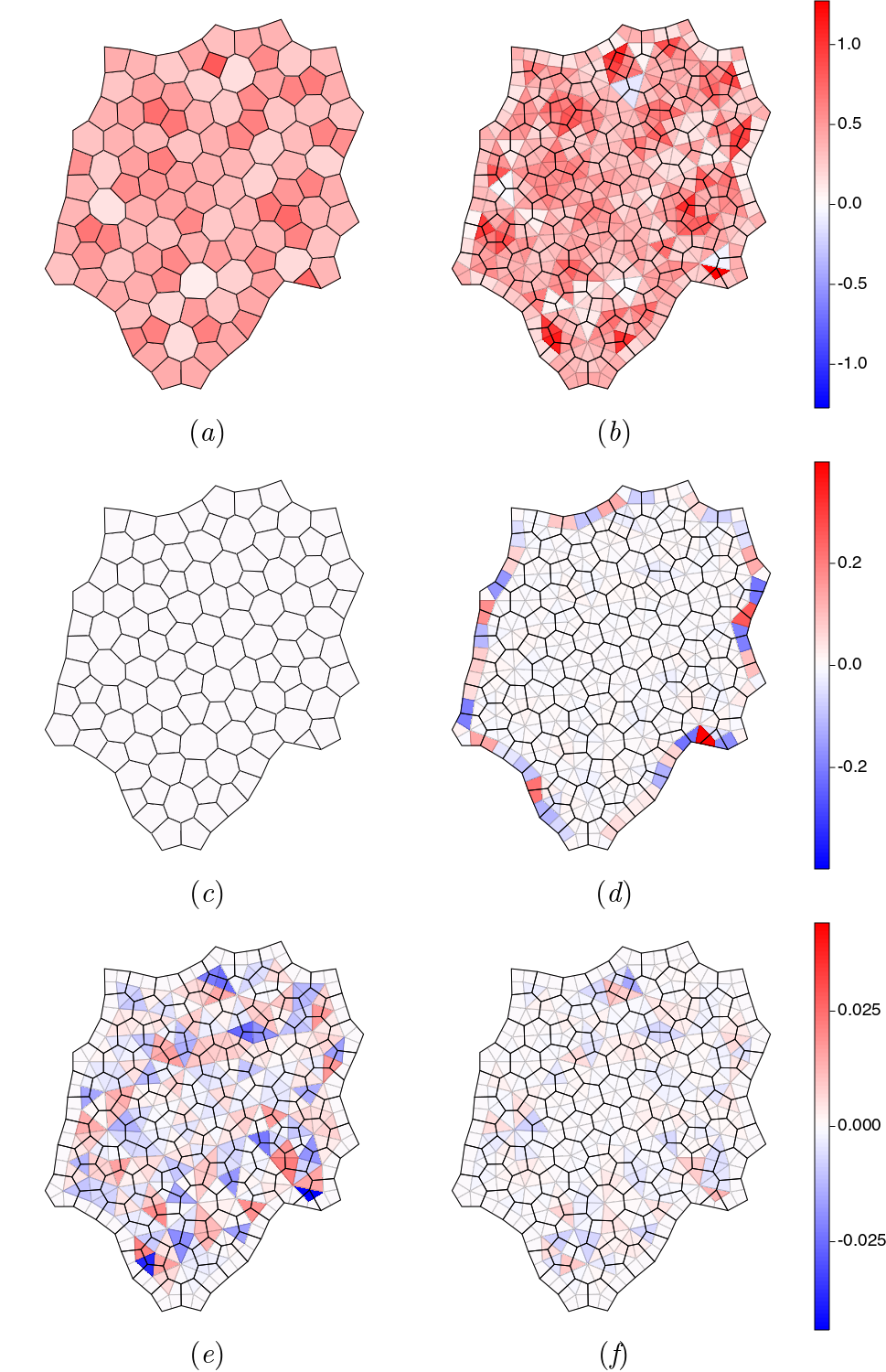}
    \caption{For the equilibrium state shown in Figure~\ref{fig:geometry}a, we show (a)  $-\tfrac{1}{2} \mathrm{div}^c \,\mathbf{h}$, giving the effective pressure in cells, (b) $-\tfrac{1}{2} \mathrm{div}^v \,\mathbf{h} $ giving the effective pressure over internal triangles, (c) $\mathrm{curl}^c\, \mathbf{h}$, which vanishes over cells, and (d) $\mathrm{CURL}^v \,\mathbf{h}$, giving a measure of the couple in the neighbourhood of each internal vertex, accounting for non-affine deformations.  The field in (d) is largest at the monolayer periphery; (e) shows $\mathrm{CURL}^v\,\mathbf{h}$ over internal triangles on a finer colour scale. (f) Couples $\mathcal{C}_k$ computed using (\ref{eq:tricouple}) under an affine approximation.  {\color{black}The three} colour bars (right) apply to (a,b), (c,d) and (e,f) {\color{black}respectively}.}
    \label{fig:divcurl}
\end{figure}

The vertices $\mathbf{h}_j$ of the rotated force network were then used to evaluate predictions of the model.   We evaluated $-\tfrac{1}{2} \{ \mathrm{div}^c \,\mathbf{h} \}_i$ and confirmed that it recovered $P_{\mathrm{eff}i}$ in (\ref{eq:peff}) (Figure~\ref{fig:divcurl}a), while $-\tfrac{1}{2}\mathrm{div}^v \,\mathbf{h}$ gives the corresponding effective pressure partitioned over triangles (Figure~\ref{fig:divcurl}b).  The fields show similar patterns over large scales.
We validated the prediction that $\mathrm{curl}^c \,\mathbf{h}=0$ (Figure~\ref{fig:divcurl}c) but found that $\mathrm{CURL}^v \,\mathbf{h}$ typically is nonzero, being largest at the periphery (Figure~\ref{fig:divcurl}d) but heterogeneous over internal triangles (Figure~\ref{fig:divcurl}e).  For comparison, we show (in Figure~\ref{fig:divcurl}f) the predicted couple $\mathcal{C}_k$ (\ref{eq:tricouple}). The patterns are distinct and differ in magnitude.  The couple acting over trijunctions and at the monolayer periphery (\ref{eq:peffk}), evaluated using simulations that incorporate non-affine deformations, therefore differs from the couple (\ref{eq:tricouple}) predicted via an affine approximation. 

\begin{figure}
    \centering
    \includegraphics[width=\columnwidth]{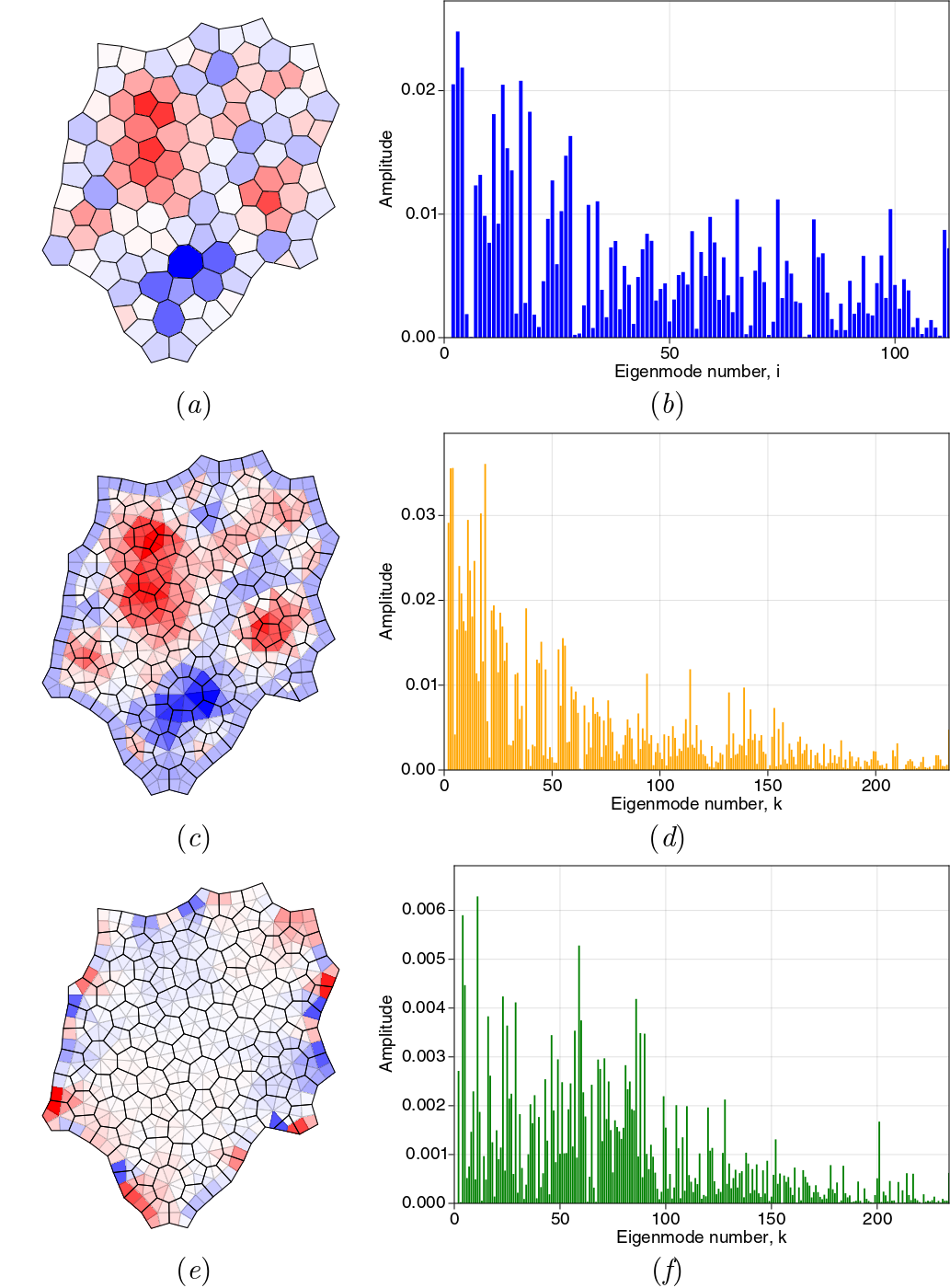}
    \caption{Potentials (a) ${\psi}^c$, (c) ${\psi}^v$ (analogues of the Airy stress function), and (e) $\check{\Psi}^v$ (analogue of the Mindlin stress function), which provide representations of the force potential over the primal network of cells. (b,d,f) show the corresponding eigenmode spectra of (a,c,e), plotting the amplitudes of coefficients in the spectral representation (\ref{eq:lapsol}).}
\label{fig:sfs}
\end{figure}

The potential $\psi^c$ (Figure~\ref{fig:sfs}a) underpins variations in the pressure field $P_{\mathrm{eff}i}$, and is built from eigenmodes of $\mathsf{L}_\mathcal{F}$.  Its spectrum shows contributions from a high proportion of modes (Figure~\ref{fig:sfs}b).  Because high-order eigenmodes are localised around defects (Figure~\ref{fig:laps}), the spectrum demonstrates the influence of these small-scale structures on the global stress field.  The potential $\psi^v$ (Figure~\ref{fig:sfs}c) shows a similar distribution over the monolayer (except for a peripheral layer) and also has a broad spectrum  (Figure~\ref{fig:sfs}d). These potentials are smoother functions than $P_\mathrm{eff}$, the latter being a second derivative of the former, but show that the pressure fields are not harmonic (unlike classical linear elasticity).  
$\check{\Psi}^v$ is largest at the periphery (Figure~\ref{fig:sfs}e), reflecting the structure of $\mathrm{CURL}^v\,\mathbf{h}$, and also has a broad spectrum (Figure~\ref{fig:sfs}f).

\begin{figure}
    \centering
    \includegraphics[width=\columnwidth]{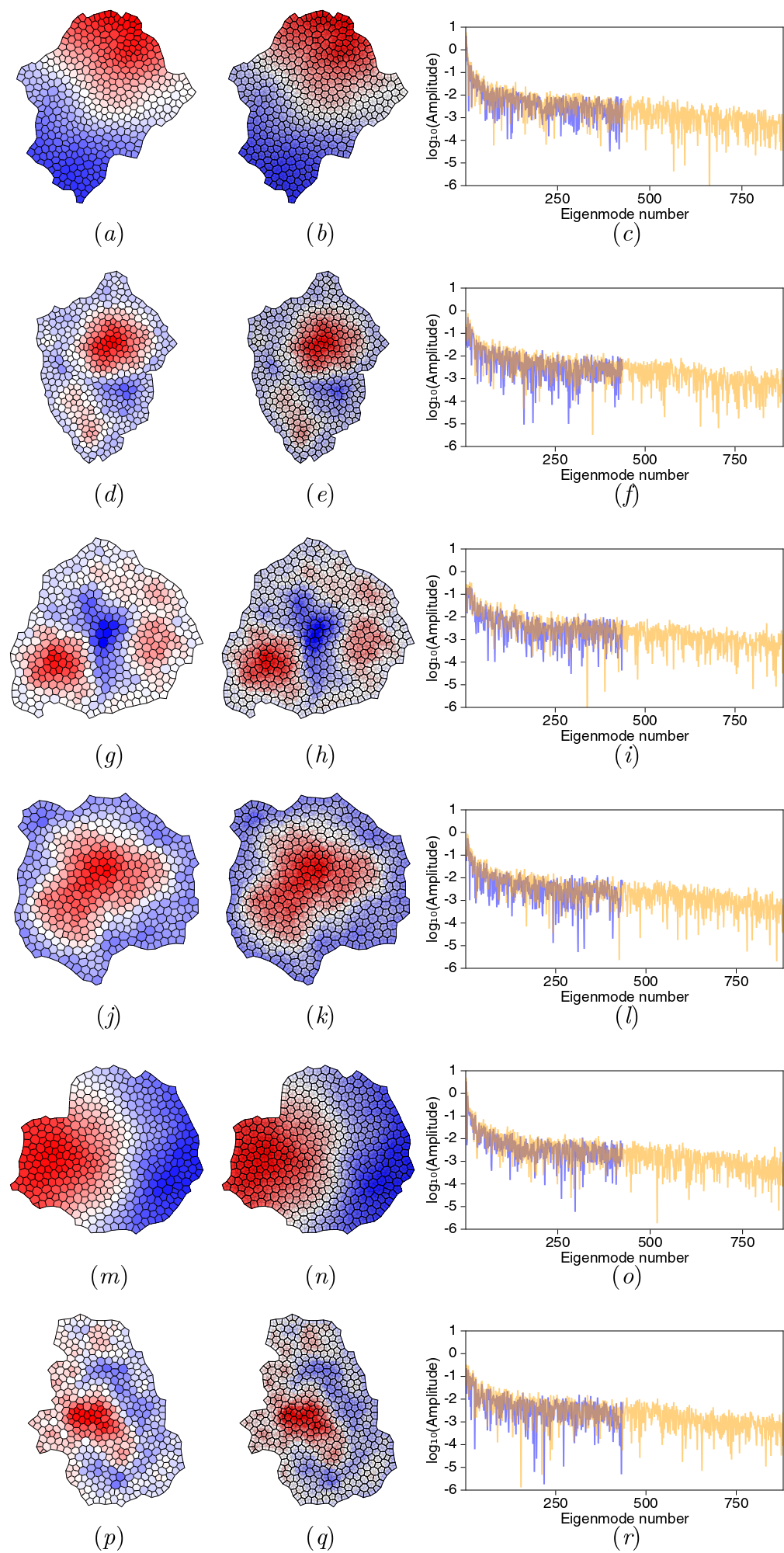}
    \caption{Airy stress functions, $\psi^c$ (a, d, g, j, m) and $\psi^v$ (b, e, h, k, n), in 6 realisations of localised monolayers with corresponding eigenmode spectra (c, f, i, l, o) with amplitudes on a $\log_{10}$ scale ($\psi_c$ spectra blue; $\psi_v$ spectra orange).}
    \label{fig:airymontage}
\end{figure}

Finally, a set of {\color{black} equilibrium} monolayers were generated using a random cell division algorithm.  $\psi^c$ and $\psi^v$ were evaluated to illustrate a range of possible patterns (Figure~\ref{fig:airymontage}).  In each case, the two potentials resemble each other at the macroscopic scale, but also show heterogeneities at the smallest scales.  Eigenvalue spectra are typically broad, although low-order modes can have prominent contributions.  These examples show diverse patterns of residual stress within the equilibrium monolayers. 

\section{Discussion}
\label{sec:disc}

Continuum mechanical models are widely used to describe biological tissues, and do so successfully over length-scales that are large in comparison to a tissue's internal heterogeneities.  However at scales comparable to individual cells, the inherent granularity of the tissue becomes evident.  The vertex model \cite{weliky1990, nagai2001, farhadifar2007, fletcher2014, alt2017} is one of a class of discrete models of tissue mechanics that resolves stresses at the level of individual cells, exploiting the natural partitioning of space that they provide.  This offers immediate advantages in modelling growth processes, by allowing cell division, expansion and rearrangement to be represented explicitly, and capturing growth-induced residual stresses.  Likewise, explicit representation of individiual cells facilitates the description of subcellular processes (such as the cell cycle, or cell signalling) and enables direct comparison with images.  Continuum models rely on assumed strain energy functions, expressed in terms of strain invariants; in contrast, the vertex model relies on a mechanical energy defined in terms of easily measured geometric invariants (such as the area or perimeter of cells in a planar monolayer).  Despite these differences in the approach to constitutive modelling, a Cauchy stress can be defined in both instances.

In continua, it is commonly assumed that the Cauchy stress is symmetric, reflecting the absence of net torque on the smallest material elements; accordingly, stresses on material elements depend on local strains.   In a discrete model, however, the smallest elements (\hbox{e.g.} individual cells) have finite size: stresses are specified primarily by local geometric measures (deviations in cell area and perimeter from target values serve as strains) but also by spatial gradients of bulk strain, `measured' across the length of an individual element.   Deformations that generate appropriate {\color{black}in-plane} bending may thereby generate torques on tissue elements that are accommodated by so-called couple stresses.  The present study is the first (to our knowledge) to address this feature in models of multicellular tissues, by evaluating the couple stress associated with the traditional vertex model.  Our study of passive torques in epithelia is distinct from that of Yamamoto \hbox{et al.} \cite{yamamoto2020}, who consider active cortical torque generation in a vertex model using a `disk-shaft' mechanism.

The monolayers addressed here are deliberately simple: they are mechanically passive and confluent \cite{kim2021}, and do not demonstrate fluctuations, motility or slippage of adjacent junctions \cite{nestor2022}.  The strain energy that we chose to investigate (\ref{eq:energy}) passes a number of basic tests.  Imposing forces at cell vertices is sufficient to ensure zero net force on each cell.  This is demonstrated by closed loops in the plane of rotated forces (Figure~\ref{fig:forcenet}b); similar networks are used in granular flows \cite{ramola2017} and suspensions \cite{edens2021}, and help visualise heterogeneous stress patterns.  The model also ensures zero net torque on individual cells, and a cell force-stress tensor (\ref{eq:varstress}) that is symmetric (which we validated numerically in Figure~\ref{fig:divcurl}c).  The stress of a cell can be constructed by summing contributions from individual vertices (or equivalently, from individual edges).  These contributions can be repartitioned to evaluate the stress over the triangulation connecting cell centres.  Here, in contrast, we find that the force-stress tensor is asymmetric (Figure~\ref{fig:divcurl}d,e), implying that a torque is exerted in the neighbourhood of each tricellular junction.  A couple stress must be incorporated in order to accommodate the torque.

Bearing in mind the simplicity of the constitutive model (\ref{eq:energy}), it is perhaps unsurprising that analogies between the vertex model and continuum models are imperfect.  This can be anticipated, given the significance of non-affine deformations in fibre networks \cite{chandran2006affine}, which can limit the accuracy of continuum approximations that assume affine deformations \cite{stracuzzi2022}.  One route to couple stress is to consider the rotational contribution to the vector force potential (\ref{eq:hmu}), that generates asymmetries in stress tensors defined over triangles spanning cell vertices.  An alternative route considers the couple stress as a quantity that is energy-conjugate to {\color{black}in-plane} bending deformations, represented by the curvature vector $\boldsymbol{\kappa}$.  Both routes indicate the existence of couple stress, as a vector defined over edges and links that has zero curl over cells, but the predictions differ in detail (Figure~\ref{fig:divcurl}e,f).  The former route evaluates torques directly in terms of computed forces, accounting for non-affine deformations.  The latter route rests on an affine assumption.  The vector $\mathfrak{m}$ (\ref{eq:cstress}) is nevertheless of interest as it suggests a direct interpretation of torques at vertices arising from pressure differences between the three cells neighbouring a junction, acting over edges of different lengths, creating a net moment (\ref{eq:tricouple}).  It also suggests that couple stresses are intrinsically connected to spatial disorder, given that perfectly symmetric cells do not show the area-response to {\color{black}in-plane} bending of asymmetric cells in (\ref{eq:la}b).  However it underpredicts the overall torque (Figure~\ref{fig:divcurl}e,f), likely because local equilibration at vertices leads to deformations not captured in an affine approximation \cite{chandran2006affine, stracuzzi2022}.  

An array of confluent polygonal cells provides a natural unstructured mesh on which to perform computations.  The machinery for pursuing such calculations is provided by discrete calculus, combining tools of algebraic topology \cite{desbrun2005, grady2010} with mimetic finite differences \cite{daveiga2014, lipnikov2014}.  Incidence matrices $\mathsf{A}$ and $\mathsf{B}$ encode topological relationships between cell vertices, edges and faces, and the equivalent relationships over the dual triangulation.  When combined with appropriate metric information, they can be used to construct discrete differential operators.  By respecting the need to preserve exact sequences, a full set of operators can be identified, including positive-semi-definite discrete Laplacians defined over the primal (cell) and dual (triangular) networks (Table~\ref{table1}, Figure~\ref{fig:operators}), having eigenmodes (Figure~\ref{fig:laps}) with which potentials can be constructed.  Helmholtz--Hodge decomposition (for a monolayer with no holes) enables a vector field defined over cell edges (namely, a force potential built from the forces acting on cell vertices) to be represented in terms of scalar potentials on each network.  Thereby, we recover the discrete analogues of the Airy stress function of traditional 2D continuum elasticity, and the additional function introduced by Mindlin to describe couple stresses (Figure~\ref{fig:sfs}).  In general, the functions derived over networks of cells are distinct from those derived over the dual triangulation, although they share large-scale features (Figures~\ref{fig:sfs}a,b \& \ref{fig:airymontage}). Broad eigenvalue spectra (Figures~\ref{fig:sfs}, \ref{fig:airymontage}) implicate small-scale features near topological defects (Figure~\ref{fig:laps}) in overall stress patterns.

With this framework in place, we return to a question raised previously \cite{jensen2020}, namely the consequence of neglecting torque balance in computational implementation of the vertex model.  If couple stresses are assumed not to exist, so that all stresses are symmetric (over cells and over the dual triangulation), then cell edges and links between cell should, in principle, be orthogonal.  Indeed, a stronger condition was identified (that vertices sit at the orthocentre of the triangle formed by their neighbours \cite{jensen2020}), which suppresses shearing deformations and is typically violated in real monolayers.  Invoking couple stress relaxes the orthogonality (and orthocentricity) constraint, but reveals distributions of torques across the monolayer.  These are largest at the monolayer periphery (boundary-layer features being characteristic of couple-stress materials \cite{toupin1964}) but distributed also across the interior of the monolayer (Figure~\ref{fig:divcurl}c,e).  Couples are relatively weak in comparison to other stresses but they highlight tricellular junctions as sites where asymmetries in cell packing may be detected by mechanosensitive processes.

In the present analysis of force- and couple-stress in cellular monolayers,
we have considered only systems at equilibrium, and not have not accounted for transient viscous effects or neighbour exchanges.  However this assessment of the vertex model demonstrates its utility in crossing scales from cell to tissue.  Identifying the Laplacians of the cellular network opens the door for spectral methods to investigate global patterns of stress a tissue in a systematic way.  Force chains within cell monolayers or associated with cells embedded in matrix \cite{mann2019} are an interesting target for investigation, as they may provide a mechanism for long-range mechanical signalling.  More generally, this study also highlights a requirement to recognise that disordered multicellular tissues may need to be modelled at the macroscopic level as couple-stress materials, with boundary-layer effects (Figure~\ref{fig:divcurl}d) and torques at tricellular junctions (Figure~\ref{fig:divcurl}e) emerging as essential features.  







\backmatter





\section*{Declarations}

For the purpose of open access, the authors have applied a Creative Commons Attribution (CC BY) licence to any Author Accepted Manuscript version arising.

\begin{itemize}
\item Funding: This work was supported by Biotechnology and Biological Sciences Research Council (BB/T001984/1) and The Leverhulme Trust (RPG-2021-394).
\item The authors declare no competing interests.
\item Ethics approval: n/a
\item Consent to participate: n/a
\item Consent for publication: n/a
\item Availability of data and materials: All code used to generate data reported here is provided at \cite{Revell_VertexModel_jl_2022}.
\item Code availability: VertexModel.jl https://github.com/chris-revell/VertexModel
\item Authors' contributions: OEJ drafted and edited the manuscript.  CKR performed computations, created figures and edited the manuscript.
\end{itemize}

\begin{appendices}

\section{Couple stress in 2D continua}
\label{sec:couple}

In two dimensions, a continuous simply-connected couple-stress material in plane strain can be characterised by a force-stress tensor $\boldsymbol{\sigma}$ (having zero divergence, ensuring force balance) and couple-stress vector $\boldsymbol{\mu}$, with the antisymmetric component of $\boldsymbol{\sigma}$ expressed as a curl of $\boldsymbol{\mu}$ (ensuring torque balance).  There are three independent components of the symmetric component of force stress $\boldsymbol{\sigma}^{(s)}\equiv \tfrac{1}{2}( \boldsymbol{\sigma}+ \boldsymbol{\sigma}^\top)$ and two of $\boldsymbol{\mu}$, constrained by two (scalar) force balances and a torque balance.  These constraints are satisfied by
expressing $\boldsymbol{\sigma}$ and $\boldsymbol{\mu}$ in terms of two potentials, the Airy stress function $\psi(\mathbf{x})$ plus a second stress function $\Psi(\mathbf{x})$ described by Mindlin \cite{mindlin1962}, such that \cite{hadjesfandiari2011}
\begin{equation}
\sigma_{pq}=\varepsilon_{pr}\partial_r (\varepsilon_{qs} \partial_s \psi - \partial_q \Psi), \quad \mu_p=-\varepsilon_{pq}\partial_q \Psi,
\label{eq:mind}
\end{equation}
ensuring that $\partial_p \sigma_{pq}=0$ ($\mathrm{div}\,\boldsymbol{\sigma}=\mathbf{0}$) and $\partial_p \mu_p=0$ ($\mathrm{div}\,\boldsymbol{\mu}={0}$).  Here $\boldsymbol{\varepsilon}$ is the 2D Levi-Civita tensor representing a clockwise $\pi/2$ rotation; note that $\mathrm{curl}\,\phi=\boldsymbol{\varepsilon}\nabla \phi$ for a scalar $\phi$ and $\mathrm{curl}\,\mathbf{b}=\nabla\cdot(\boldsymbol{\varepsilon} \mathbf{b})$ for a vector $\mathbf{b}$, so that $\mathrm{curl} \, \mathrm{curl}\, \phi \equiv - \nabla^2 \phi$, allowing (\ref{eq:mind}) to be written as (\ref{eq:curlcurl}).  In Cartesians, (\ref{eq:mind}) becomes
\begin{gather*}
\sigma_{xx}=\partial_y^2 \psi-\partial_x \partial_y \Psi, \quad
\sigma_{yy}=\partial_x^2 \psi+\partial_x \partial_y \Psi, \\
\sigma_{xy}=-\partial_x\partial_y \psi- \partial_y^2 \Psi, \quad
\sigma_{yx}=-\partial_x\partial_y \psi+\partial_x^2 \Psi
\end{gather*}
with $\mu_y=\partial_x \Psi$ and $-\mu_x=\partial_y \Psi$.
This formulation makes minimal constitutive assumptions beyond material continuity, except that the condition $\mathrm{div}\,\boldsymbol{\mu}={0}$ is a compatibility condition for an isotropic linearly elastic material rather than an equilibrium condition \cite{hadjesfandiari2011}.
The Airy stress function $\psi$ determines the isotropic component of the force-stress via $\mathrm{Tr}(\boldsymbol{\sigma})=\nabla^2 \psi$, while the Mindlin stress function $\Psi$ determines the antisymmetric force stress via the torque balance $\boldsymbol{\sigma}^{(a)}\equiv \tfrac{1}{2} (\boldsymbol{\sigma}-\boldsymbol{\sigma}^\top)=-\tfrac{1}{2} \boldsymbol{\varepsilon} \nabla^2 \Psi = - \tfrac{1}{2} \boldsymbol{\varepsilon} (\partial_x \mu_y - \partial_y \mu_x)$.   The force stress can be decomposed into isotropic, antisymmetric and symmetric-deviatoric parts as
$\boldsymbol{\sigma}=\tfrac{1}{2}\mathsf{I} \nabla^2 \psi -\tfrac{1}{2}\boldsymbol{\varepsilon} \nabla^2 \Psi  + {\boldsymbol{\sigma}}^{(s)}$, 
where $\mathrm{Tr}({\boldsymbol{\sigma}}^{(s)})=0$.  ${\boldsymbol{\sigma}}^{(s)}$ has real eigenvalues $\pm \lambda$, with $\lambda\geq 0$ measuring shear, which depends on both $\psi$ and $\Psi$ via
\begin{multline}
\lambda^2=\left[\tfrac{1}{2}(\partial_y^2 -\partial_x^2)\psi-\partial_x \partial_y \Psi \right]^2 \\ + \left[ \partial_x\partial_y \psi +\tfrac{1}{2}(\partial_y^2 -\partial_x^2) \Psi \right]^2.
\end{multline}
Writing $\mathbf{h}=-\nabla \psi-\boldsymbol{\mu}$ implies that $\boldsymbol{\sigma}=-\tfrac{1}{2} \mathsf{I} \nabla\cdot \mathbf{h} + \tfrac{1}{2} \boldsymbol{\varepsilon} \, \mathrm{curl}\, \mathbf{h} + {\boldsymbol{\sigma}}^{(s)}$, and
\begin{equation}
\label{eq:sigmah}
\sigma_{pq}=\varepsilon_{pr}\partial_r (-\varepsilon_{qs} h_s).
\end{equation}
In this sense, $\mathbf{h}$ can be regarded as a vector potential for the force stress, and $\psi$ and $\Psi$ can be regarded as  scalar potentials of $\mathbf{h}$ in a Helmholtz decomposition ($\mathbf{h}$ being the sum of a gradient of $\psi$ and a curl of $\Psi$).

The gradient of a smooth deformation $\mathbf{u}(\mathbf{x})$ can be decomposed into $\mathsf{E}+\mathsf{W}$, where $\mathsf{E}=\mathsf{E}^\top\equiv \tfrac{1}{2}(\nabla\mathbf{u} + \nabla\mathbf{u}^\top)$ represents strain and $\mathsf{W}\equiv \tfrac{1}{2}(\nabla\mathbf{u} - \nabla\mathbf{u}^\top)=\boldsymbol{\varepsilon} \omega$ is a rotation, where $\omega\equiv \tfrac{1}{2}\nabla\cdot(\boldsymbol{\varepsilon}\mathbf{u})$.  Likewise, $\mathsf{M}\equiv (\nabla\otimes \nabla) \mathbf{u}$ can be deomposed as $\mathsf{M}=\nabla \mathsf{E}+\nabla \mathsf{W}$.  $\mathsf{M}$ is symmetric in its first two arguments, while contracting over them gives
\begin{equation}
\mathsf{I}:\nabla \mathsf{W} = \tfrac{1}{2} (\nabla^2 \mathbf{u}-\nabla(\nabla\cdot\mathbf{u}))\equiv  -2\boldsymbol{\kappa},
\end{equation}
which defines a curvature vector $\boldsymbol{\kappa} =\tfrac{1}{2} \boldsymbol{\varepsilon} \nabla \omega$ \cite{hadjesfandiari2011}.  The corresponding principle of virtual work for a continuous couple-stress material occupying a volume $\mathcal{V}$ can then be written \cite{hadjesfandiari2011}
\begin{multline}
\int_{\mathcal{V}} \left(\boldsymbol{\sigma}^{(s)}: \delta \mathsf{E} -2 \boldsymbol{\mu}\cdot \delta \boldsymbol{\kappa} \right)\,\mathrm{d}V \\ =\int_{\partial\mathcal{V}} \left( \boldsymbol{\tau}\cdot \delta \mathbf{u} + {m} \delta \omega \right)\,\mathrm{d}S,
\label{eq:vw}
\end{multline}
showing that the curvature vector is energy-conjugate to the couple-stress vector.  Here $\boldsymbol{\tau}=\mathbf{n}\cdot\boldsymbol{\sigma}$ is a force traction at a surface with unit normal $\mathbf{n}$, and ${m}=\mathbf{n}\cdot \boldsymbol{\varepsilon} \boldsymbol{\mu}$ is a couple traction.

\section{Discrete operators}
\label{sec:operators}

The discrete analogues of 2D differential operators are defined over primal and dual networks and act on variables defined on vertices, edges, and faces of each \cite{grady2010, daveiga2014}.  These are summarised in Fig.~\ref{fig:operators}, which shows the {primary operators} on the cell network ($\mathrm{grad}^v$, $\mathrm{curl}^v$, $\mathrm{curl}^c$ and $\mathrm{div}^c$) and on the dual network ($\mathrm{grad}^c$, $\mathrm{CURL}^c$, $\mathrm{CURL}^v$ and $\mathrm{div}^v$).

\subsection{Operators on the primary network}

We define vector spaces $\mathcal{V}$, $\mathcal{E}$, $\mathcal{F}$ of fields defined on vertices, edges and faces, with associated {inner products} $[\cdot,\cdot]_\mathcal{V}$, $[\cdot,\cdot]_\mathcal{E}$, $[\cdot,\cdot]_\mathcal{F}$ represented by matrices $\mathsf{M}^{\mathcal{V}}=\mathrm{diag}(E_k)$, $\mathsf{M}^{\mathcal{E}}=\mathrm{diag}(\mathsf{I} F_j)$, $\mathsf{M}^{\mathcal{F}}=\mathrm{diag}(A_i)$, where $\mathsf{I}$ is the $2\times 2$ identity.  $\mathsf{M}^{\mathcal{E}}$ is influenced by both edges and links via (\ref{eq:trapezium}). 
Thus
\begin{subequations}
\label{eq:inprod}
\begin{align}
[\phi,\psi]_{\mathcal{V}}& \equiv  {\textstyle \sum_{k, k'}} \phi_k M^{\mathcal{V}}_{k k'} \psi_{k'}={\textstyle \sum_{k}}  E_k \phi_k \psi_{k}, \\
[\mathbf{u},\mathbf{v} ]_{\mathcal{E}}& \equiv {\textstyle \sum_{j, j'}} \mathbf{u}_j^\top \mathsf{M}^{\mathcal{E}}_{j j'} \mathbf{v}_{j'} = {\textstyle \sum_{j}} F_j \mathbf{u}_j \cdot \mathbf{v}_{j}, \\
[f,g]_{\mathcal{F}}& \equiv  {\textstyle \sum_{i, i'}} f_i M^{\mathcal{F}}_{i i'} g_{i'} =  {\textstyle \sum_{i}}  A_i f_i g_{i},
\end{align}
\end{subequations}
for any $\phi, \psi \in \mathcal{V}$, $\mathbf{u}, \mathbf{v} \in \mathcal{E}$, $f, g \in \mathcal{F}$.

The `primary operators' (in the terminology of \cite{daveiga2014}) over cells are $\mathrm{grad}^v:\mathcal{V}\rightarrow \mathcal{E}$, $\mathrm{curl}^v:\mathcal{V}\rightarrow \mathcal{E}$, $\mathrm{curl}^c:\mathcal{E}\rightarrow \mathcal{F}$ and $\mathrm{div}^c:\mathcal{E}\rightarrow \mathcal{F}$, and are defined as in Table~\ref{table1}.
$\mathrm{curl}^c$ and $\mathrm{div}^c$ mimic the integrals arising in Stokes and divergence theorems.  In matrix form,
\begin{subequations}
\label{eq:primdef}
\begin{align}
 \mathrm{grad}^v\, & = (\mathsf{N}^{\mathcal{E}})^{-1}\mathsf{A} \mathsf{N}^{\mathcal{V}}, &
 \mathrm{curl}^c\, &=(\mathsf{N}^{\mathcal{F}})^{-1} \mathsf{B} \mathsf{N}^{\mathcal{E}}, \\
 \mathrm{curl}^v  &= (\tilde{\mathsf{N}}^{\mathcal{E}})^{-1}\mathsf{A} \mathsf{N}^{\mathcal{V}}, &
 \mathrm{div}^c\,&= (\mathsf{N}^{\mathcal{F}})^{-1} \mathsf{B}\tilde{\mathsf{N}}^{\mathcal{E}},
\end{align}
\end{subequations}
where $\mathsf{N}^{\mathcal{V}}=\mathsf{I}$, $\mathsf{N}^{\mathcal{E}}=\mathrm{diag}(\mathbf{t}_j^\top)$, $\mathsf{N}^{\mathcal{F}}=\mathrm{diag}(A_i)$ and $\tilde{\mathsf{N}}^{\mathcal{E}}=\mathrm{diag}(-(\boldsymbol{\epsilon}_i\mathbf{t}_j)^\top)$ (so $(\mathsf{N}^{\mathcal{E}})^{-1}=\mathrm{diag}(\mathbf{t}_j/t_j^2)$, $(\tilde{\mathsf{N}}^{\mathcal{E}})^{-1}=\mathrm{diag}((\boldsymbol{\epsilon}_k\mathbf{t}_j)/t_j^2)$).   The topological relationship $\mathsf{B}\mathsf{A}=\mathsf{0}$ ensures that $\mathrm{curl}^c \circ \, \mathrm{grad}^v=0$ and $\mathrm{div}^c \circ \, \mathrm{curl}^v=0$.  These exact sequences (de Rahm complexes) and definitions (\ref{eq:primdef}) can be represented using the commutative diagrams
\begin{equation*}
\begin{CD}
\mathcal{V}  @>{\mathrm{grad}^v}>> \mathcal{E}^\parallel @>{\mathrm{curl}^c}>> \mathcal{F} \\
@VV{\mathsf{N}^{\mathcal{V}}}V @VV{\mathsf{N}^{\mathcal{E}}}V @VV{\mathsf{N}^{\mathcal{F}}}V \\
\mathcal{V} @>\mathsf{A}>> \mathcal{E}^\parallel @>\mathsf{B}>>  \mathcal{F}
\end{CD}
\end{equation*} and
\begin{equation*}
\begin{CD}
\mathcal{V}  @>{\mathrm{curl}^v}>> \mathcal{E}^\perp @>{-\mathrm{div}^c}>> \mathcal{F} \\
@VV{\mathsf{N}^{\mathcal{V}}}V @VV{\tilde{\mathsf{N}}^{\mathcal{E}}}V @VV{\mathsf{N}^{\mathcal{F}}}V \\
\mathcal{V} @>\mathsf{A}>> \mathcal{E}^\perp @>\mathsf{B}>>  \mathcal{F}
\end{CD}
\end{equation*}
where $\mathcal{E}$ is the direct sum of spaces of vectors parallel and perpendicular to edges $\mathcal{E}^\parallel\oplus\mathcal{E}^\perp$.

Derived operators (denoted with tildes, following \cite{daveiga2014}) are defined as adjoints of the primary operators under the inner products (\ref{eq:inprod}), satisfying
\begin{subequations}
\label{eq:adj}
\begin{align}
[\mathrm{grad}^v\,\phi, \mathbf{b}]_{\mathcal{E}}&=[\phi,-\widetilde{\mathrm{div}}^v\, \mathbf{b}]_{\mathcal{V}}, \\
[\mathrm{curl}^c\,\mathbf{b}, f]_{\mathcal{F}}&=[\mathbf{b},\widetilde{\mathrm{curl}}^c f]_{\mathcal{E}},\\
[\mathrm{curl}^v\,\phi, \mathbf{b}]_{\mathcal{E}}&=[\phi,\widetilde{\mathrm{curl}}^v\, \mathbf{b}]_{\mathcal{V}}, \\
[-\mathrm{div}^c\, \mathbf{b}, f]_{\mathcal{F}}&=[\mathbf{b},\widetilde{\mathrm{grad}}^c f]_{\mathcal{E}},
\end{align}
\end{subequations}
for any $\phi \in \mathcal{V}$, $\mathbf{b} \in \mathcal{E}$, $f \in \mathcal{F}$.  It follows from (\ref{eq:inprod}, \ref{eq:adj}) that the derived operators have the matrix representations
\begin{subequations}
\label{eq:firstformulation}
\begin{align}
\widetilde{\mathrm{grad}}^c&=-(\mathsf{M}^{\mathcal{E}})^{-1} \mathrm{div}^{c\top} \mathsf{M}^{\mathcal{F}}, \\
\widetilde{\mathrm{curl}}^c&=(\mathsf{M}^{\mathcal{E}})^{-1} \mathrm{curl}^{c\top} \mathsf{M}^{\mathcal{F}}, \\
\widetilde{\mathrm{curl}}^v&=(\mathsf{M}^{\mathcal{V}})^{-1} \mathrm{curl}^{v\top} \mathsf{M}^{\mathcal{E}}, \\
\widetilde{\mathrm{div}}^v&=-(\mathsf{M}^{\mathcal{V}})^{-1}\mathrm{grad}^{v\top} \mathsf{M}^{\mathcal{E}}.
\end{align}
\end{subequations}
These relationships are summarised in the upper half of Figure~\ref{fig:operators}.
Under (\ref{eq:firstformulation}), $\widetilde{\mathrm{div}}^v \circ \,\widetilde{\mathrm{curl}}^c=0$ and $\widetilde{\mathrm{curl}}^v \circ \, \widetilde{\mathrm{grad}}^c=0$ are both satisfied exactly: for example,
$-\widetilde{\mathrm{div}}^v \circ \widetilde{\mathrm{curl}}^c=(\mathsf{M}^{\mathcal{V}})^{-1} (\mathrm{grad}^v)^\top \mathsf{M}^{\mathcal{E}} \circ (\mathsf{M}^{\mathcal{E}})^{-1} (\mathrm{curl}^c)^\top \mathsf{M}^{\mathcal{F}} 
=(\mathsf{M}^{\mathcal{V}})^{-1} ((\mathsf{N}^{\mathcal{E}})^{-1} \mathsf{A}\mathsf{N}^{\mathcal{V}})^\top ((\mathsf{N}^{\mathcal{F}})^{-1} \mathsf{B} \mathsf{N}^{\mathcal{E}} )^\top \mathsf{M}^{\mathcal{F}} 
=(\mathsf{M}^{\mathcal{V}})^{-1} (\mathsf{N}^{\mathcal{V}})^\top \mathsf{A}^\top \mathsf{B}^\top ((\mathsf{N}^{\mathcal{F}})^{-1})^\top   \mathsf{M}^{\mathcal{F}}$,
which vanishes because $(\mathsf{B} \mathsf{A})^\top=\mathsf{0}$.  
The derived operators are given in Table~\ref{table1}.

By specifying fields appropriately in (\ref{eq:adj}), we can use (\ref{eq:adj}) to write, for any $\phi\in \mathcal{V}$, $\mathbf{b}\in\mathcal{E}$ and $f\in\mathcal{V}$,
\begin{subequations}
\begin{align}
0\leq [\mathrm{grad}^v\,\phi, \mathrm{grad}^v\,\phi]_{\mathcal{E}}&=[\phi,\mathsf{L}_\mathcal{V}
\phi]_{\mathcal{V}}, \\
0\leq [\mathrm{curl}^v\,\phi, \mathrm{curl}^v\,\phi]_{\mathcal{E}}&=[\phi,
\mathsf{L}_\mathcal{V} \phi]_{\mathcal{V}}, \\
0 \leq [\widetilde{\mathrm{grad}}^c\, f, \widetilde{\mathrm{grad}}^c\, f]_{\mathcal{E}}&=[f,\mathsf{L}_\mathcal{F}
f]_{\mathcal{F}}, \\
0\leq [\widetilde{\mathrm{curl}}^c\,f, \widetilde{\mathrm{curl}}^c\,f ]_{\mathcal{E}}&=[f,\mathsf{L}_\mathcal{F}
f]_{\mathcal{F}},
\end{align}
\label{eq:prods}
\end{subequations}
where $\mathsf{L}_\mathcal{V}=-\widetilde{\mathrm{div}}^v\,\circ \mathrm{grad}^v=\widetilde{\mathrm{curl}}^v\,\circ \mathrm{curl}^v$ and $\mathsf{L}_{\mathcal{F}}= -\mathrm{div}^c \circ \widetilde{\mathrm{grad}}^c =  \mathrm{curl}^c\circ\widetilde{\mathrm{curl}}^c$ (see Figure~\ref{fig:operators}), showing that these scalar Laplacians (given in matrix form in (\ref{eq:lapmat})) are positive-semi-definite.  (Regularisation of $\mathsf{L}_\mathcal{C}$ and $\mathsf{L}_\mathcal{F}$ at the monolayer periphery, shown in (\ref{eq:lapmat}b), is explained in Sec.~\ref{sec:invlap} below.)


We can use (\ref{eq:adj}) to obtain the orthogonality relations
\begin{equation}
\label{eq:orth}
[\mathrm{grad}^v\,\phi, \widetilde{\mathrm{curl}}^c f]_{\mathcal{E}}=0, \quad
[\mathrm{curl}^v\, \phi , \widetilde{\mathrm{grad}}^c f]_{\mathcal{E}}=0,
\end{equation}
which hold for any functions $\phi\in \mathcal{V}$ and $f\in \mathcal{F}$.  These results rely on $\mathsf{B}\mathsf{A}=\mathsf{0}$, rather than geometric orthogonality, and they underpin the Helmholtz decomposition (\ref{eq:h1}).

We also identify two Helmholtzians (closely related to Hodge Laplacians $\mathsf{A} \mathsf{A}^\top+\mathsf{B}^\top \mathsf{B}$) acting on $\mathcal{E}$,
\begin{align*}
\mathcal{H}_1&=\widetilde{\mathrm{curl}}^c\circ \mathrm{curl}^c-\widetilde{\mathrm{grad}}^v\circ \mathrm{div}^v, \\
\mathcal{H}_2&={\mathrm{curl}}^v\circ\widetilde{\mathrm{curl}}^v-{\mathrm{grad}}^c\circ \widetilde{\mathrm{div}}^c,
\end{align*}
which sit in the spaces spanned by $\hat{\mathbf{t}}_j \otimes \hat{\mathbf{t}}_j$ and $(\boldsymbol{\epsilon}_k\hat{\mathbf{t}}_j) \otimes (\boldsymbol{\epsilon}_k \hat{\mathbf{t}}_j)$ respectively.  We assume that $\mathrm{dim}(\mathrm{ker} (\mathcal{H}_1))=0$ and $\mathrm{dim}(\mathrm{ker} (\mathcal{H}_2))=0$, ensuring that there are no additional harmonic contributions to the decomposition (\ref{eq:h1}).  A necessary condition is that the domain has no holes \cite{lim2020}.  In this case, $\mathrm{ker}(\mathrm{curl}^c)=\mathrm{im}(\mathrm{grad}^v)$, $\mathrm{ker}(\mathrm{div}^c)=\mathrm{im}(\mathrm{curl}^v)$, with analogous results for derived operators \cite{daveiga2014}, ensuring sequences in Figure~\ref{fig:operators} are exact.

\subsection{Operators on the dual network}

Primary operators on the dual network are defined as follows:
\begin{equation*}
\begin{CD}
\mathcal{T}  @<{\mathrm{CURL}^v}<< \mathcal{L}^\parallel @<{\mathrm{grad}^c}<< \mathcal{C} \\
@VV{\mathsf{N}^{\mathcal{T}}}V @VV{\mathsf{N}^{\mathcal{L}}}V @VV{\mathsf{N}^{\mathcal{C}}}V \\
\mathcal{T} @<\mathsf{A}^\top << \mathcal{L}^\parallel @<\mathsf{B}^\top <<  \mathcal{C}
\end{CD}
\end{equation*}
and
\begin{equation*}
\begin{CD}
\mathcal{T}  @<{-\mathrm{div}^v}<< \mathcal{L}^\perp @<{\mathrm{CURL}^c}<< \mathcal{C} \\
@VV{\mathsf{N}^{\mathcal{T}}}V @VV{\tilde{\mathsf{N}}^{\mathcal{L}}}V @VV{\mathsf{N}^{\mathcal{C}}}V \\
\mathcal{T} @<\mathsf{A}^\top<< \mathcal{L}^\perp @<\mathsf{B}^\top<<  \mathcal{C}
\end{CD}
\end{equation*}
where $\mathcal{L}$ is the direct sum of spaces of vectors parallel and perpendicular to links, $\mathcal{L}^\parallel\oplus\mathcal{L}^\perp$.
Here $\mathsf{N}^{\mathcal{C}}=\mathsf{I}$, $\mathsf{N}^{\mathcal{L}}=\mathrm{diag}(\mathbf{T}_j^\top )$, $\tilde{\mathsf{N}}^{\mathcal{L}}=\mathrm{diag}(-(\boldsymbol{\epsilon}_k\mathbf{T}_j)^\top)$, $\mathsf{N}^{\mathcal{T}}=\mathrm{diag}(E_k)$.  $\mathcal{T}$, $\mathcal{L}$ and $\mathcal{C}$ are vector spaces of fields defined over triangles, links and cell centres.  We note the isomorphisms $\mathcal{T}\simeq \mathcal{V}$, $\mathcal{L}\simeq \mathcal{E}$, $\mathcal{C}\simeq \mathcal{F}$.   Derived operators are defined using the inner products with metrics $\mathsf{M}^{\mathcal{T}}=\mathsf{M}^{\mathcal{V}}$, $\mathsf{M}^{\mathcal{L}}=\mathsf{M}^{\mathcal{E}}$, $\mathsf{M}^{\mathcal{C}}=\mathsf{M}^{\mathcal{F}}$, via sequences shown in the lower half of Figure~\ref{fig:operators}.
This leads to operators given in Table~\ref{table1}, with actions illustrated in the lower half of Figure~\ref{fig:operators}, and the Helmholtz decomposition given in (\ref{eq:h2}).

\begin{figure*}
    \centering
    \includegraphics[width=0.89\textwidth]{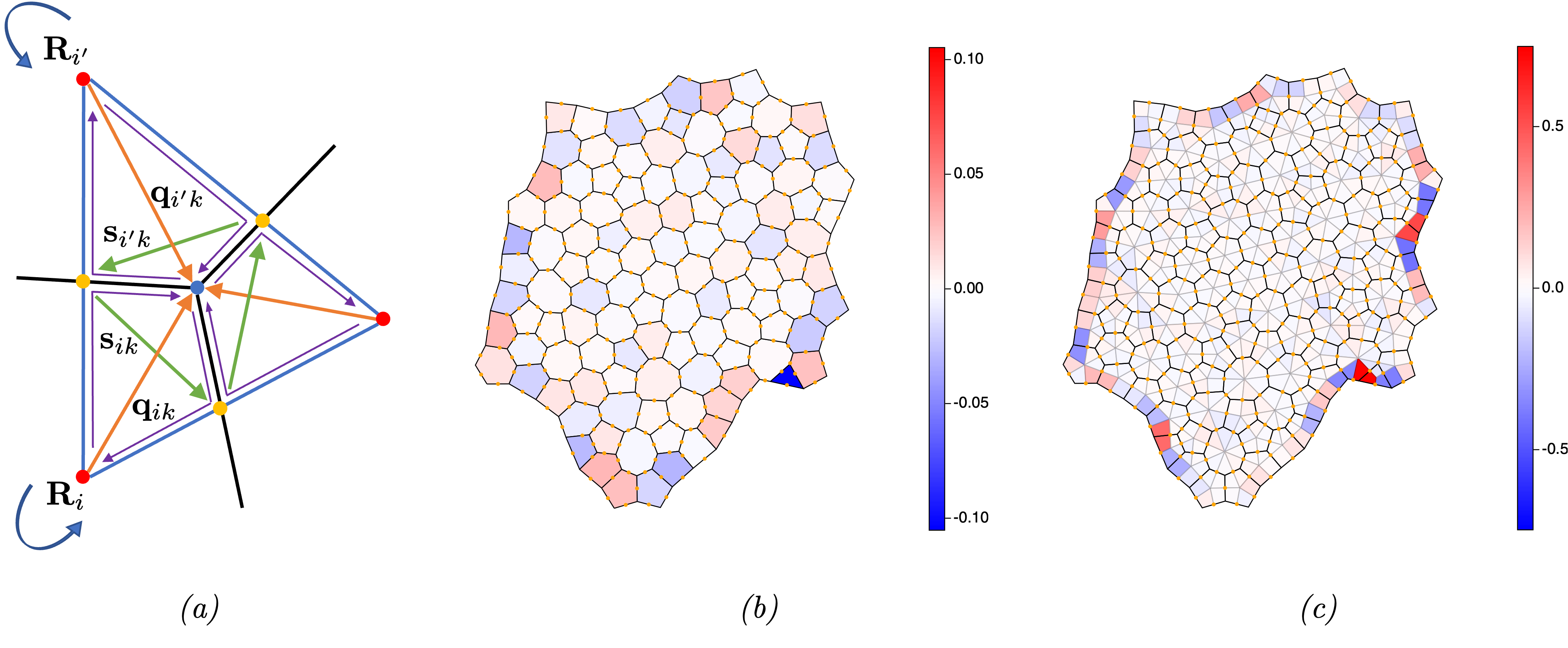}
    \caption{(a) A diagram illustrating how a curl around the edges of a triangle centred on vertex $k$ (blue dot) can be modified to the path taken by thin purple lines, which reduce to integrals along spokes $\mathbf{q}_{ik}$ (orange), as expressed in (\ref{eq:periphop}).  Green lines $\mathbf{s}_{ik}$ connect intersections $\mathbf{m}_j$ between edges and links (yellow dots), so that $\mathbf{q}_{ik}$ and $\mathbf{s}_{ik}$ span $m$-kite $ik$.
    (b) $\mathrm{curl}^c\,\mathbf{m}$ and (c) $\mathrm{CURL}^v\,\mathbf{m}$ provide maps of cell and triangle asymmetry over a monolayer. Intersections $\mathbf{m}$ of edges and links are shown with orange dots. }
    \label{fig:peripheralkite}
\end{figure*}

Additional steps are necessary to accommodate faces of the dual network at the monolayer periphery, which are not complete triangles.  Writing $\mathbf{q}_{ik}=C_{ik}(\mathbf{r}_k-\mathbf{R}_i)$ as the spoke connecting cell centre $\mathbf{R}_i$ to adjacent vertex $\mathbf{r}_k$, we can alternatively write
\begin{subequations}
\label{eq:periphop}
\begin{align}
    \left\{\mathrm{CURL}^v\,\mathbf{b}\right\}_k&=-\frac{1}{E_k} {\textstyle \sum_{i,j}} B_{ij} A_{jk} \mathbf{q}_{ik}\cdot\mathbf{b}_j,\\
        \left\{\mathrm{div}^v\,\mathrm{b}_j\right\}_k&=-\frac{1}{E_k} {\textstyle \sum_{i,j}} B_{ij} A_{jk} \boldsymbol{\epsilon}_i \mathbf{q}_{ik}\cdot\mathbf{b}_j.
\end{align}
A geometric interpretation for an internal triangle is provided in Figure~\ref{fig:peripheralkite}(a): the path around the boundary of a triangle can be reformulated into paths around the component kites (doubling back on all internal edges); since $\mathbf{b}_j$ is uniform over each link, the paths are equivalent to running up and down each spoke.  Evaluation over spokes allows definition of adjoint operators as
\begin{align*}
    \left\{\widetilde{\mathrm{CURL}}^v\,\phi \right\}_j&=-\frac{1}{F_j} {\textstyle \sum_{i,k}} B_{ij} A_{jk} \mathbf{q}_{ik} \phi_k,\\
        \left\{ \widetilde{\mathrm{grad}}^v\,\phi\right\}_j&=\frac{1}{F_j} {\textstyle \sum_{i,k}} B_{ij} A_{jk} \boldsymbol{\epsilon}_i \mathbf{q}_{ik}\phi_k.
\end{align*}
\end{subequations}

Eq.~(\ref{eq:periphop}) is advantageous as it holds for peripheral faces of the dual network, made from just one or two kites.  This enables the discrete analogues of the divergence and Stokes theorems to be stated.  Transforming sums over cells [triangles] to sums over edges [links], making use of derived operators defined in Table~\ref{table1} and (\ref{eq:periphop}a), it follows for a vector field $\mathbf{h}$ that
\begin{subequations}
\label{eq:stokes}
\begin{align}
    [\mathbbm{1}^c,-\mathrm{div}^c \mathbf{h}]_\mathcal{F}=& [\widetilde{\mathrm{grad}}^c\mathbbm{1}^c,\mathbf{h}]_\mathcal{E}={\textstyle \sum_j} \boldsymbol{\epsilon}_i \mathbf{t}^p_j \cdot \mathbf{h}_j^p, \\
[\mathbbm{1}^v,-\mathrm{div}^v \mathbf{h}]_\mathcal{V}=& [\widetilde{\mathrm{grad}}^v\mathbbm{1}^v,\mathbf{h}]_\mathcal{L}={\textstyle \sum_j} \boldsymbol{\epsilon}_i\mathbf{t}^p_j \cdot \mathbf{h}_j^p \\
    [\mathbbm{1}^c,\mathrm{curl}^c \mathbf{h}]_\mathcal{F}=& [\widetilde{\mathrm{curl}}^c\mathbbm{1}^c,\mathbf{h}]_\mathcal{E}={\textstyle \sum_j} \mathbf{t}^p_j \cdot \mathbf{h}_j^p, \\
[\mathbbm{1}^v,\mathrm{CURL}^v \mathbf{h}]_\mathcal{V}=& [\widetilde{\mathrm{CURL}}^v\mathbbm{1}^v,\mathbf{h}]_\mathcal{L} =-{\textstyle \sum_j} \mathbf{t}^p_j \cdot \mathbf{h}_j^p
\end{align}
\end{subequations}
where $\mathbf{t}_j^p$ are edges at the periphery of the monolayer.

\subsection{Validation}

To validate (\ref{eq:periphop}), we define $\mathbf{m}_j$ to be the intersection of edge $\mathbf{t}_j$ and link $\mathbf{T}_j$.  For all internal intersections,
$\mathbf{m}_j=\hat{\mathbf{r}}_k+F_j^{-1}{(\hat{\mathbf{R}}_i-\hat{\mathbf{r}}_k)\cdot\boldsymbol{\epsilon}_k \mathbf{T}_j}{} \mathbf{t}_j$,
where $\hat{\mathbf{r}}_k=\tfrac{1}{2}(\overline{A}_{jk}-A_{jk})\mathbf{r}_k$ and $\hat{\mathbf{R}}_i=\tfrac{1}{2}(\overline{B}_{ij}-B_{ij})\mathbf{R}_i$ lie at the base of $\mathbf{t}_j$ and $\mathbf{T}_j$ respectively.  In general, $\mathbf{m}_j$ lies close to, but is distinct from, the edge centroid $\mathbf{c}_j$.   However we define $\mathbf{m}_j=\mathbf{c}_j^p$ on peripheral edges.  (We must therefore distinguish $m$-kites with vertices $\mathbf{m}_j$, shown in Figure~\ref{fig:peripheralkite}a, from $c$-kites with vertices $\mathbf{c}_j$, shown in Figure~\ref{fig:geometry}b).  Writing $\mathbf{s}_{ik}=\sum_j B_{ij}\mathbf{m}_j A_{jk}$, then $\mathbf{s}_{ik}$ and $\mathbf{q}_{ik}$ span each $m$-kite within the monolayer.  $\sum_k C_{ik}\mathbf{s}_{ik}\otimes \mathbf{q}_{ik}$ gathers $m$-kites into cells, while $\sum_i C_{ik}\mathbf{s}_{ik}\otimes \mathbf{q}_{ik}$ gathers $m$-kites into faces of the dual network (including the single and double kites at the periphery).  Making use of $\mathbf{s}_{ik}\otimes\mathbf{q}_{ik}=K_{ik}^m\boldsymbol{\varepsilon}+\mathsf{F}_k$, where $K_{ik}^m$ is the area of the $m$-kite spanned by $\mathbf{s}_{ik}$ and $\mathbf{q}_{ik}$ and $\mathsf{F}_{ik}$ is a (symmetric) fabric tensor that measures asymmetries in kite shape \cite{jensen2020}, it follows that $\boldsymbol{\varepsilon} \mathbf{q}_{ik}\cdot\mathbf{s}_{ik}=2K_{ik}^m$ and, because $E_k=\sum_i K_{ik}^m$,
\begin{equation}
    \mathrm{div}^c\,\mathbf{m}=2\mathbbm{1}^c, \quad \mathrm{div}^v\,\mathbf{m}=2\mathbbm{1}^v
    \label{eq:m}
\end{equation}
over the entire monolayer, providing a useful test for computations.   
In contrast, $\mathrm{curl}^c\,\mathbf{m}$ and $\mathrm{CURL}^v\,\mathbf{m}$
provide a map of the asymmetry of each face of the primal and dual networks (Figure~\ref{fig:peripheralkite}b,c).  This is largest in the peripheral cells, but is distributed across the monolayer.  The opposite orientations of cells and triangles accounts for the opposite signs of curls, for example in the neighbourhood of the two triangular peripheral cells.

\subsection{Inverting Laplacians}
\label{sec:invlap}

We can partition vertices into peripheral and interior vertices, edges into peripheral, border and interior edges, and cells into border and interior cells, so that \cite{jensen2020}
\begin{align*}
    \mathsf{H}=&\left(\begin{matrix}\mathsf{H}^b & \mathsf{0} \\ \mathsf{0} & \mathsf{H}^i \end{matrix}\right), \quad
    \mathsf{T}_e=\left(\begin{matrix}\mathsf{T}_e^p & \mathsf{0} & \mathsf{0} \\ \mathsf{0} & \mathsf{T}_e^b & 0 \\ \mathsf{0} & \mathsf{0} & \mathsf{T}_e^i \end{matrix}\right), \\
    \mathsf{B}=&\left(\begin{matrix}\mathsf{B}^{bp} & \mathsf{B}^{bb} & \mathsf{B}^{bi} \\ \mathsf{0} &\mathsf{0} & \mathsf{B}^{ii} \end{matrix}\right).
\end{align*}
Then the unregularised Laplacian $\mathsf{L}_\mathcal{F}=\mathsf{H}^{-1}\mathsf{B}\mathsf{T}_e \mathsf{B}^\top$ becomes
\begin{subequations}
\begin{equation}
    \mathsf{L}_\mathcal{F}
    =\left(\begin{matrix}
    (\mathsf{H}^{b}){^{-1}} \mathsf{B}^{bp}\mathsf{T}_e^p (\mathsf{B}^{bp})^\top & \mathsf{0} \\ \mathsf{0} & \mathsf{0}
    \end{matrix}\right)+ \mathsf{L}^r_{\mathcal{F}}
\label{eq:lf}
    \end{equation}
where
    \begin{equation}
\mathsf{L}^r_{\mathcal{F}}=\left(\begin{matrix}
    \mathsf{L}_\mathcal{F}^{rbb} & (\mathsf{H}^{b}){^{-1}} \mathsf{B}^{bi}\mathsf{T}_e^i (\mathsf{B}^{ii})^\top \\  (\mathsf{H}^{i}){^{-1}} \mathsf{B}^{ii}\mathsf{T}_e^i (\mathsf{B}^{bi})^\top &  \mathsf{L}_\mathcal{F}^{rii}
    \end{matrix}\right),
\end{equation}
\end{subequations}
with $\mathsf{L}_\mathcal{F}^{rbb}=(\mathsf{H}^{b}){^{-1}}( \mathsf{B}^{bb}\mathsf{T}_e^b (\mathsf{B}^{bb})^\top+ \mathsf{B}^{bi}\mathsf{T}_e^i (\mathsf{B}^{bi})^\top)$ and $\mathsf{L}_\mathcal{F}^{rii}=(\mathsf{H}^{i}){^{-1}} \mathsf{B}^{ii}\mathsf{T}_e^i (\mathsf{B}^{ii})^\top$.
The first contribution to $\mathsf{L}_\mathcal{F}$ in (\ref{eq:lf}) involves `orphan' links between border cell centres and peripheral edge centroids via $\mathsf{B}^{bp}$, whereas the remaining `regularised' component of the Laplacian, $\mathsf{L}^r_\mathcal{F}$, involves links between adjacent border cells via $\mathsf{B}^{bb}$, links between border and interior cells via $\mathsf{B}^{bi}$ and links between interior cells via $B^{ii}$.  Since all such links lie between adjacent cells, $\mathsf{L}^r_\mathcal{F}$ satisfies $\mathsf{L}^r_\mathcal{F} \mathbbm{1}^c=\mathsf{0}$, and we use this in building potentials.  $\mathsf{L}_\mathcal{V}^r$ is most easily evaluated by setting all elements of $\mathsf{T}_e^p$ to zero when computing $\mathsf{H}^{-1}\mathsf{B}\mathsf{T}_e \mathsf{B}^\top$, as in (\ref{eq:lapmat}b).  $\mathsf{L}^r_\mathcal{C}$ can be evaluated similarly, setting to zero all peripheral elements of $\mathsf{T}_l^{-1}$.

From (\ref{eq:prods}a), $\mathsf{L}_{\mathcal{V}}$ satisfies
$[\phi,\mathsf{L}_{\mathcal{V}}\phi]_{\mathcal{V}}\equiv \phi^\top \mathsf{E} \mathsf{L}_{\mathcal{V}}\phi\geq 0$.  Noting that $\mathsf{E}^\top=\mathsf{E}$ and $\mathsf{L}_{\mathcal{V}}=\mathsf{E}^{-1} \mathsf{L}_{\mathcal{V}}^\top \mathsf{E}$, self-adjointness is demonstrated from
$[\phi,\mathsf{L}_{\mathcal{V}}\psi]_{\mathcal{V}}
    =\phi^\top \mathsf{E} \mathsf{L}_{\mathcal{V}} \psi
    =\psi^\top  \mathsf{L}_{\mathcal{V}}^\top \mathsf{E} \phi
    =\psi^\top \mathsf{E} \mathsf{E}^{-1} \mathsf{L}_{\mathcal{V}}^\top \mathsf{E} \phi =\psi^\top \mathsf{E} \mathsf{L}_{\mathcal{V}}  \phi
  =[\psi , \mathsf{L}_{\mathcal{V}} \phi]_{\mathcal{V}}
    =[\mathsf{L}_{\mathcal{V}} \phi,\psi]_{\mathcal{V}}$.
Let $\lambda_k^\mathcal{V}$ and ${e}_k^\mathcal{V}$ ($k=1,\dots,\lambda_{N_v}$) be the eigenvalues and eigenvectors of $\mathsf{L}_\mathcal{V}$.  Then
$[{e}_p^\mathcal{V},\mathsf{L}_{\mathcal{V}} {e}_q^\mathcal{V}]_{\mathcal{V}}
    =\lambda_q^\mathcal{V}[{e}_p^\mathcal{V},{e}_q^\mathcal{V}]_{\mathcal{V}}
=[\mathsf{L}_{\mathcal{V}}{e}_p^\mathcal{V}, {e}_q^\mathcal{V}]_{\mathcal{V}}
   =\lambda_p^\mathcal{V}[{e}_p^\mathcal{V}, {e}_q^\mathcal{V}]_{\mathcal{V}}$,
and so $(\lambda_q^\mathcal{V}-\lambda_p^\mathcal{V})[{e}_p^\mathcal{V},{e}_q^\mathcal{V}]_{\mathcal{V}}=0$, demonstrating orthogonality (under the inner product) of eigenvectors having distinct eigenvalues.  Writing $\phi=\sum_k c_k{e}_k^\mathcal{V}$ and projecting onto ${e}_p^\mathcal{V}$, it follows that
\begin{equation}
    \phi={\textstyle\sum_k} \frac{[{e}_k^\mathcal{V},\phi]_{\mathcal{V}}}{[{e}_k^\mathcal{V},{e}_k^\mathcal{V}]_{\mathcal{V}}} {e}_k^\mathcal{V}.
    \label{eq:inversion}
\end{equation}

Now $\lambda_1^\mathcal{V}=0$ and ${e}_1^\mathcal{V}=\mathbbm{1}^v$, with remaining eigenvalues being positive.  To invert $\mathsf{L}_{\mathcal{V}}\phi =g$, we define $\overline{g}\equiv [\mathbbm{1}^v,g]_{\mathcal{V}}\mathbbm{1}^v/[\mathbbm{1}^v,\mathbbm{1}^v]_\mathcal{V}$ and set $\breve{g}\equiv g-\overline{g}$, so that $[\mathbbm{1}^v,\breve{g}]_{\mathcal{V}}=0$.  Then write $\phi=\overline{\phi}+\breve{\phi}$ where
\begin{equation}
    \mathsf{L}_{\mathcal{V}}\overline{\phi}=\overline{g}, \quad
    \mathsf{L}_{\mathcal{V}}\breve{\phi}=\breve{g}.
\label{eq:gbre}
\end{equation}
(\ref{eq:gbre}a) has the solution $\overline{\phi}=([\mathbbm{1}^v,g]_{\mathcal{V}}/[\mathbbm{1}^v,\mathbbm{1}^v]_\mathcal{V}) \overline{\phi}_0$ where $\mathsf{L}_{\mathcal{V}}\overline{\phi}_0=\mathbbm{1}^v$, while (\ref{eq:gbre}b) has forcing satisfying the solvability condition $[\mathbbm{1}^v, \breve{g}]_\mathcal{V}=0$.  This enables its solution to be expressed in terms of the remaining eigenmodes, to obtain     $\phi=\overline{\phi}+\breve{\phi}$ where
\begin{equation}
    \overline{\phi}=\frac{[\mathbbm{1}^v,g]_{\mathcal{V}}}{[\mathbbm{1}^v,\mathbbm{1}^v]_\mathcal{V}}
    \overline{\phi}_0,\quad \breve{\phi}=\sum_{k=2}^{N_v} \frac{1}{\lambda_k} \frac{[{e}_k^\mathcal{V},\breve{g}]_{\mathcal{V}}}{[{e}_k^\mathcal{V},{e}_k^\mathcal{V}]_{\mathcal{V}}} {e}_k^\mathcal{V}.
\label{eq:lapsol}
\end{equation}
Similar arguments follow for $\mathsf{L}_{\mathcal{T}}$ under $[\cdot, \cdot]_\mathcal{V}$, and $\mathsf{L}_{\mathcal{C}}^r$ and $\mathsf{L}_{\mathcal{F}}^r$ under $[\cdot, \cdot]_\mathcal{F}$.  The $\mathbf{m}$ field, with uniform divergence but non-zero curl (Figure~\ref{fig:peripheralkite}b,c), allows evaluation of $\psi^c$ and $\check{\psi}^v$ via (\ref{eq:pot1}a, \ref{eq:divvcurlv}a).

\section{Representations of stress}
\label{sec:str}

\subsection{Microscopic force stress}
\label{sec:micro}

The conservation principle (\ref{eq:outerbc}) makes it possible to consider how $A_i \boldsymbol{\sigma}^c_i$ or $E_k \boldsymbol{\sigma}^v_k$ are built from their component parts.  Accordingly, we define the microscopic force stress as
\begin{subequations}
    \label{eq:microstresses}
\begin{gather}
    A_i \tilde{\boldsymbol{\sigma}}_i^c={\textstyle{\cup_j}}  B_{ij} ( \mathbf{t}_j\otimes \mathbf{h}_j) \boldsymbol{\epsilon}_i, \\
    E_k \tilde{\boldsymbol{\sigma}}_k^v={\textstyle{\cup_j}} A_{jk} (\mathbf{T}_j \otimes \mathbf{h}_j) \boldsymbol{\epsilon}_k.
\end{gather}
\end{subequations}
Here $\mathsf{B}$ [$\mathsf{A}$] attributes each edge [link] component to a neighbouring cell [triangle] face but maintains it as a distinct entity from the other edge [link] contributions.  It follows immediately that $\mathbf{n}_{ij}\cdot \tilde{\boldsymbol{\sigma}}_i^c=0$ for each edge $j$ of cell $i$ and $\mathbf{N}_{jk} \cdot \tilde{\boldsymbol{\sigma}}_k^v=0$ for each link $j$ of triangle $k$, ensuring that
$\mathrm{div}^c\,\tilde{\boldsymbol{\sigma}}_i^c=\mathbf{0}$ and $\mathrm{div}^v \, \tilde{\boldsymbol{\sigma}}_k^v=\mathbf{0}$.


{We decompose $\mathbf{h}_j$ into components along $\mathbf{t}_j$ and $\boldsymbol{\epsilon_{i}}\mathbf{t}_j$, so that
\begin{align}
A_i \tilde{\boldsymbol{\sigma}}_i^c & ={\textstyle{\cup_j}}  B_{ij} \big[ \hat{\mathbf{t}}_j\otimes \hat{\mathbf{t}}_j (\mathbf{h}_j\cdot \mathbf{t}_j) \nonumber \\ & \qquad \qquad + \hat{\mathbf{t}}_j\otimes (\boldsymbol{\epsilon}_i \hat{\mathbf{t}}_j)(\mathbf{h}_j\cdot (\boldsymbol{\epsilon}_i \mathbf{t}_j))\big] \boldsymbol{\epsilon}_i \nonumber \\
& ={\textstyle{\cup_j}} \big[ \hat{\mathbf{t}}_j\otimes \hat{\mathbf{n}}_{ij} (\mathbf{h}_j\cdot \mathbf{t}_j) \nonumber \\ & \qquad \qquad +  B_{ij}  \hat{\mathbf{t}}_j\otimes  \hat{\mathbf{t}}_j(\mathbf{h}_j\cdot (\boldsymbol{\epsilon}_i \mathbf{t}_j))\big],
\label{eq:cstrh}
\end{align}
where hats denote unit vectors.  Then $A_i\mathrm{Tr}( \tilde{\boldsymbol{\sigma}}_i^c) ={\textstyle{\sum_{j'}}}  B_{ij'}  \mathbf{h}_{j'}\cdot (\boldsymbol{\epsilon}_i \mathbf{t}_{j'})$ so that the deviatoric microscopic force stress becomes
\begin{align*}
A_i \tilde{\boldsymbol{\sigma}}_i^{cD}  &={\textstyle{\cup_j}}   \big[ \hat{\mathbf{t}}_j\otimes \hat{\mathbf{n}}_{ij} (\mathbf{h}_j\cdot \mathbf{t}_j) \nonumber \\   + B_{ij} \hat{\mathbf{t}}_j\otimes  \hat{\mathbf{t}}_j& (\mathbf{h}_j\cdot (\boldsymbol{\epsilon}_i \mathbf{t}_j))\big] -\tfrac{1}{2}\mathsf{I} {\textstyle{\sum_{j'}}}  B_{ij'}  \mathbf{h}_{j'}\cdot (\boldsymbol{\epsilon}_i \mathbf{t}_{j'}).
\end{align*}
The final term involves $\sum_j$ rather than $\cup_j$, ensuring that $\mathrm{div}^c\,\tilde{\boldsymbol{\sigma}}_i^{cD}=\mathbf{0}$.  $\tilde{\boldsymbol{\sigma}}_i^{cD}$ has symmetric component
\begin{multline}
A_i {\tilde{\boldsymbol{\sigma}}}_i^{c(s)}  ={\textstyle{\cup_j}}   \big[\tfrac{1}{2} (\hat{\mathbf{t}}_j\otimes \hat{\mathbf{n}}_{ij}+\hat{\mathbf{n}}_{ij}\otimes \hat{\mathbf{t}}_{j}) (\mathbf{h}_j\cdot \mathbf{t}_j) \nonumber \\
+ B_{ij} \hat{\mathbf{t}}_j\otimes  \hat{\mathbf{t}}_j(\mathbf{h}_j\cdot (\boldsymbol{\epsilon}_i \mathbf{t}_j))\big] 
-\tfrac{1}{2}\mathsf{I} {\textstyle{\sum_{j'}}}  B_{ij'}  \mathbf{h}_{j'}\cdot (\boldsymbol{\epsilon}_i \mathbf{t}_{j'})
\end{multline}
and antisymmetric component
\begin{align}
A_i {\tilde{\boldsymbol{\sigma}}}_i^{c(a)} &={\textstyle{\cup_j}}  \tfrac{1}{2} (\hat{\mathbf{t}}_j\otimes \hat{\mathbf{n}}_{ij}-\hat{\mathbf{n}}_{ij}\otimes \hat{\mathbf{t}}_{j}) (\mathbf{h}_j\cdot \mathbf{t}_j) \nonumber \\ 
&=\tfrac{1}{2} \boldsymbol{\epsilon}_i {\textstyle{\cup_j}}   B_{ij}  (\mathbf{h}_j\cdot \mathbf{t}_j),
\label{eq:edgecouple}
\end{align}
where we have used $\hat{\mathbf{t}}_j\otimes \hat{\mathbf{n}}_{ij}-\hat{\mathbf{n}}_{ij}\otimes \hat{\mathbf{t}}_{j}\equiv \boldsymbol{\epsilon}_iB_{ij}$ (consider its action on a vector $\alpha \hat{\mathbf{t}_j}+\beta \hat{\mathbf{n}}_{ij}$).
We can therefore interpret $\mathbf{h}_j\cdot \mathbf{t}_j$ in (\ref{eq:edgecouple}) as a torque exerted on each edge of the cell.  Analogous expressions to (\ref{eq:cstrh}-\ref{eq:edgecouple}) follow immediately for $\tilde{\boldsymbol{\sigma}}_k$, after projecting $\mathbf{h}_j$ onto  $\mathbf{T}_j$ and $\boldsymbol{\epsilon}_k \mathbf{T}_j$.}

The cell and triangle force-stresses can be recovered from microstresses by replacing $\cup_j$ with $\sum_j$ in (\ref{eq:microstresses}), as in \cite{jensen2020}, to give (\ref{eq:stresses}).

We can also draw a distinction between $\boldsymbol{\varsigma}_i^{c}$ in (\ref{eq:cstress}), the force stress integrated over cell $i$, and the corresponding {microscopic} cell stress
\begin{equation}
\label{eq:microcell}
\tilde{\boldsymbol{\varsigma}}_i^{c}=\mathcal{P}_i'\mathsf{I}+ \frac{\mathcal{T}_i'}{A_i'} \cup_j \overline{B}_{ij} \mathbf{t}_j'\otimes \hat{\mathbf{t}}_j',
\end{equation}
which retains edge-to-edge variation rather than averaging over the perimeter.   This stress has zero divergence, because evaluating $\mathrm{div}^c \,\tilde{\boldsymbol{\varsigma}}_i^{c}$ includes $\sum_j \mathbf{n}_{ij}'$ summed around a closed loop, which vanishes, and $\sum_j \mathbf{n}_{ij}' \cdot \mathbf{t}_j' \otimes \hat{\mathbf{t}}_j'$, which also vanishes as $\mathbf{n}_{ij}'\cdot \mathbf{t}_j'=0$ along each edge.  This ensures zero net force on each cell as in (\ref{eq:for}), and as illustrated by closed polygons in the $\mathbf{h}$-plane in Figure~\ref{fig:forcenet}(b).  

\subsection{Discrete potentials}
\label{sec:sdp}

Here we briefly verify the conditions in (\ref{eq:tra}).  Contracting expressions in (\ref{eq:dfp}), $\mathrm{Tr}(\boldsymbol{\sigma}^c)=\mathrm{curl}^c \circ (-\boldsymbol{\epsilon}_i \mathbf{h}) = -\mathsf{L}_{\mathcal{F}} \psi^c$ (note that $\mathrm{curl}^c\otimes \mathrm{curl}^v =0 $ because of geometry and $\mathrm{curl}^c \otimes \mathrm{grad}^v =0$ because of topology).   Likewise $\mathrm{Tr}(\boldsymbol{\sigma}^v)=\mathrm{CURL}^v \circ(- \boldsymbol{\epsilon}_k \check{\mathbf{h}}) = -\mathsf{L}_{\mathcal{T}} \check{\psi}^v$.  From Table~\ref{table1},
\begin{multline*}
(\mathrm{CURL}^v \otimes \widetilde{\mathrm{grad}}^v \check{\Psi}^v)_{k,pq}= \\
\frac{1}{E_k} {\textstyle \sum_{j, k'} }\frac{ \{\mathbf{T}_j \otimes (\boldsymbol{\epsilon}_k \mathbf{T}_{j})\}_{pq}}{F_{j'}} A_{jk'} \check{\Psi}^v_{k'}.
\end{multline*}
Using  $\mathbf{T}_j \otimes (\boldsymbol{\epsilon}_k \mathbf{T}_{j})- \{\mathbf{T}_j \otimes (\boldsymbol{\epsilon}_k \mathbf{T}_{j})\}^\top =- \boldsymbol{\epsilon}_k T_j^2$,
we see that $(\mathrm{CURL}^v \otimes \widetilde{\mathrm{grad}}^v \check{\Psi}^v)_k^{(a)}=-\tfrac{1}{2}\boldsymbol{\epsilon}_k \mathsf{L}_{\mathcal{T}} \check{\Psi}^v$, and therefore (\ref{eq:dfp}) implies $\boldsymbol{\sigma}^{v(a)} =\tfrac{1}{2} \boldsymbol{\epsilon}_k \mathsf{L}_{\mathcal{T}} \check{\Psi}^v$.   Likewise  $\boldsymbol{\sigma}^{c(a)} =\tfrac{1}{2} \boldsymbol{\epsilon}_i \mathsf{L}_{\mathcal{F}} {\Psi}^c$.  

\section{Non-uniform deformations}
\label{sec:ap}

Here we consider how area and perimeter change under affine deformations $\mathbf{u}(\mathbf{x})$ that vary with position, over lengthscales long compared to an individual cell.  In the following, subscripts $i$, $j$, $k$ attached to $\mathbf{u}$ or its derivatives denote evaluation at $\mathbf{R}_i$, $\mathbf{c}_j$ or $\mathbf{r}_k$ respectively.   Dropping third (and higher) spatial derivatives of $\mathbf{u}$ in Taylor expansions, edges, edge lengths and normals map under the deformation to
\begin{subequations}
\begin{align}
\mathbf{t}_j&\equiv \textstyle{\sum_k} A_{jk}\mathbf{r}_k=\mathbf{t}_j'+\textstyle{\sum_k} A_{jk}\mathbf{u}(\mathbf{r}_k')\nonumber \\&=\mathbf{t}_j'+\mathbf{t}_j'\cdot(\nabla \mathbf{u})_j + \dots, \label{eq:tjp}\\
t_j&\equiv\sqrt{\mathbf{t}_j\cdot\mathbf{t}_j}=t_j'\left[ 1+\hat{\mathbf{t}}_j' \cdot \mathsf{E}_j \cdot \hat{\mathbf{t}}_j' +\dots \right],
\label{eq:lj}\\
\mathbf{n}_{ij}&\equiv -\boldsymbol{\epsilon}_i B_{ij} \mathbf{t}_{j} = \mathbf{n}_{ij}'+\mathbf{n}_{ij}'\cdot (\nabla \mathbf{u})_j+\dots\nonumber \\&
= \mathbf{n}_{ij}' \cdot \left[ \mathsf{I} + (\nabla \mathbf{u})_i+ \mathbf{v}_{ij}'\cdot  \mathsf{M}_i \right]+\dots,
\label{eq:nij}
\end{align}
\end{subequations}
where (\ref{eq:nij}) shows how the mapping is referred to an adjacent cell centre. Recall that $\nabla\mathbf{u}=\mathsf{E}+\mathsf{W}$ captures stretching and rotation and $\mathsf{M}=(\nabla\otimes\nabla) \mathbf{u}=\nabla \mathsf{E}+\nabla \mathsf{W}$ captures their gradients.   Here $\mathbf{v}_{ij}$  is the vector connecting cell centre $\mathbf{R}_i$ to an adjacent edge centroid $\mathbf{c}_j$, \hbox{i.e.} $\mathbf{v}_{ij}=\overline{B}_{ij}(\mathbf{c}_j- \mathbf{R}_i)$.  The cell centre (the vertex centroid) is also the centroid relative to edge centroids (because $\sum_k C_{ik} \mathbf{r}_k=\tfrac{1}{2} \sum_{jk} \overline{B}_{ij} \overline{A}_{jk} \mathbf{r}_k=\sum_j \overline{B}_{ij} \mathbf{c}_j$, ensuring that $\sum_j\overline{B}_{ij}\mathbf{v}_{ij}=\mathbf{0}$).  Likewise
\begin{equation}
 \mathsf{E}_j=\mathsf{E}_i+\mathbf{v}_{ij}'\cdot(\nabla \mathsf{E})_i+\dots
\label{eq:cj1}
\end{equation}
Edge centroids map to
\begin{align}
\mathbf{c}_j&=\mathbf{c}_j'+\mathbf{u}_j+\tfrac{1}{8} (\mathbf{t}_j' \cdot \nabla) (\mathbf{t}_j' \cdot \nabla) \mathbf{u} \vert_j + \dots \nonumber \\ &=
\mathbf{c}_j'+\mathbf{u}_j+\tfrac{1}{8} (\mathbf{t}_j' \otimes \mathbf{t}_j' ): \mathsf{M}_j + \dots.
\label{eq:cj}
\end{align}
Using $\mathbf{R}_i=Z_i^{-1}\sum_j \overline{B}_{ij} \mathbf{c}_j$, cell centres map to
\begin{equation}
\mathbf{R}_i=\mathbf{R}_i'+\mathbf{u}_i+\tfrac{1}{2} \mathsf{V}_i:\mathsf{M}_i+\tfrac{1}{8} \mathsf{T}_i:\mathsf{M}_i+\dots
\label{eq:ri}
\end{equation}
where  $\mathsf{V}_i\equiv Z_i^{-1}\sum_j \overline{B}_{ij} \mathbf{v}_{ij}'\otimes \mathbf{v}_{ij}'$ (arising from averaging displacements around the edges of the cell, where $\mathbf{u}_j=\mathbf{u}_i+\mathbf{v}_{ij}\cdot (\nabla \mathbf{u})_i + \tfrac{1}{2} (\mathbf{v}_{ij}\otimes \mathbf{v}_{ij}):\mathsf{M}_i+\dots$) and $\mathsf{T}_i\equiv Z_i^{-1}\sum_j \overline{B}_{ij} \mathbf{t}_j'\otimes \mathbf{t}_j'$.  Combining (\ref{eq:cj}) and (\ref{eq:ri}), links from cell centres to edge centroids map to
\begin{multline}
\mathbf{v}_{ij}=\mathbf{v}_{ij}'+\mathbf{v}_{ij}'\cdot (\nabla \mathbf{u})_i + \big[ \tfrac{1}{2} (\mathbf{v}_{ij}'\otimes \mathbf{v}_{ij}') -\tfrac{1}{2} \mathsf{V}_i \\ +\tfrac{1}{8}  (\mathbf{t}_{j}'\otimes \mathbf{t}_{j}')-\tfrac{1}{8} \mathsf{T}_i \big]:\mathsf{M}_i +\dots.
\label{eq:vij}
\end{multline}

Using (\ref{eq:cj}) and (\ref{eq:lj}), cell perimeters change according to
\begin{multline}
L_i\equiv \sum_j \overline{B}_{ij}t_j=L_i'(1+\mathsf{Q}_i : \mathsf{E}_i) \\ + \textstyle{\sum_j} \overline{B}_{ij} t_j' \hat{\mathbf{t}}_j' \cdot \left[ \mathbf{v}_{ij}' \cdot(\nabla \mathsf{E})_i \right]\cdot \hat{\mathbf{t}}_j'+\dots ,
\label{eq:xi}
\end{multline}
where $L_i' \mathsf{Q}_i\equiv\sum_j \overline{B}_{ij} \mathbf{t}_j'\otimes \hat{\mathbf{t}}_j'$.  Writing the final term as $L_i'\mathsf{X}_i
\mathrel{\lower.03em\hbox{\rlap{$\cdot$}}{:}} (\mathsf{\nabla E})_i$ reveals the 3-tensor $\mathsf{X}_i$ characterising the impact of strain gradients on cell perimeter.  
Rotation gradients $\mathsf{W}$ do not affect perimeter changes to this order.

Using (\ref{eq:nij}) and (\ref{eq:vij}), and again dropping terms beyond third derivatives, the cell area becomes
\begin{align*}
A_i\equiv &\tfrac{1}{2}\textstyle{\sum_{j}} \mathbf{n}_{ij}\cdot \mathbf{v}_{ij} \\
=&A_i'+ \textstyle{\sum_j} \big\{ \mathbf{v}_{ij}' \cdot \mathsf{E}_i \cdot\mathbf{n}_{ij}' + \tfrac{1}{2} \left[\mathbf{n}_{ij}'\cdot (\mathbf{v}_{ij}'\cdot \mathsf{M}_i ) \right] \cdot\mathbf{v}_{ij}' \\
 +\tfrac{1}{2} & \left( \left[\tfrac{1}{2} \mathbf{v}_{ij}'\otimes \mathbf{v}_{ij}' +  \tfrac{1}{8} \mathbf{t}_{j}'\otimes \mathbf{t}_{j}' -\tfrac{1}{2}\mathsf{V}_i-\tfrac{1}{8} \mathsf{T}_i\right]: \mathsf{M}_i\right) \cdot\mathbf{n}_{ij}'\big\}.
\end{align*}
The terms involving $\mathsf{V}_i$ and $\mathsf{T}_i$ vanish because $\sum_j \mathbf{n}_{ij}'=\mathbf{0}$.  Consistent with (\ref{eq:cellarea}), we note that
\begin{multline*}
\textstyle{\sum_j} \mathbf{n}_{ij}\otimes \mathbf{v}_{ij}=\sum_j \mathbf{n}_{ij}\otimes \mathbf{c}_{ij} \\ =\oint_i \hat{\mathbf{n}}\otimes \mathbf{x}  \,\mathrm{d}s=\int_i \nabla \otimes \mathbf{x} \,\mathrm{d}A=A_i \mathsf{I},
\end{multline*}
where the integral over cell $i$ is simplified by recognising that the unit normal is uniform along each edge.  By extension, integrating along edge $j$ of cell $i$,
    $\int_j \hat{\mathbf{n}}_{ij} \otimes (\mathbf{x}\otimes \mathbf{x})  \,\mathrm{d}s =\hat{\mathbf{n}}_{ij}\otimes \int_{-t_j/2}^{t_j/2} (\mathbf{c}_j+s \hat{\mathbf{t}}_j)\otimes (\mathbf{c}_j+s\hat{\mathbf{t}}_j) \,\mathrm{d}s 
    ={\mathbf{n}}_{ij}\otimes \left[ \mathbf{c}_j\otimes \mathbf{c}_j + \tfrac{1}{12} \mathbf{t}_j \otimes \mathbf{t}_j\right]$,
so that integrating around cell $i$
\begin{multline*}
  \left(  {\textstyle \sum_i}\overline{B}_{ij}\mathbf{n}_{ij} \otimes \left[ \mathbf{c}_j\otimes \mathbf{c}_j + \tfrac{1}{12} \mathbf{t}_j \otimes \mathbf{t}_j\right] \right)_{pqr}\\ =(\delta_{pr}\rho_{i,q}+\delta_{pq}\rho_{i,r})A_i
\end{multline*}
where $\boldsymbol{\rho}_i\equiv A_i^{-1}\int_i \mathbf{x}\,\mathrm{d}A$ is the area-centroid of the cell, which in general will be distinct from the vertex centroid $\mathbf{R}_i$.  It follows that
\begin{multline*}
\left(\textstyle{\sum_j} \mathbf{n}_{ij} \otimes \mathbf{v}_{ij} \otimes \mathbf{v}_{ij}\right)_{pqr} \equiv
\textstyle{\sum_j} n_{ij,p} v_{ij,q} v_{ij, r} \\
 =A_i \left[ \delta_{pq} (\rho_{i,r}-R_{i,r})+\delta_{pr} (\rho_{i,q}-R_{i,q}))\right] \\ -\tfrac{1}{12}
  {\textstyle \sum_i}\overline{B}_{ij}{n}_{ij,p}  {t}_{j,q}  {t}_{j,r} .
 \end{multline*}
Thus
$[\mathbf{n}_{ij}' 
\cdot (\mathbf{v}_{ij}'\cdot \mathsf{M}_i )]\cdot\mathbf{v}_{ij}' 
= A_i'(\boldsymbol{\rho}_i'-\mathbf{R}_i')\cdot[\nabla^2 \mathbf{u}_i+\nabla(\nabla\cdot \mathbf{u}_i)] 
-\tfrac{1}{12}   {\textstyle \sum_i}\overline{B}_{ij} \left[\mathbf{n}_{ij}'\cdot\left(\mathbf{t}_{j}'\cdot\mathsf{M}_i\right)\right]\cdot \mathbf{t}_j'
\equiv 2A_i'(\boldsymbol{\rho}_i'-\mathbf{R}_i')\cdot[\mathsf{I} :\nabla \mathsf{E}]_i 
-\tfrac{1}{12}   {\textstyle \sum_i}\overline{B}_{ij} \left[\mathbf{n}_{ij}'\cdot\left(\mathbf{t}_{j}'\cdot\mathsf{M}_i\right)\right]\cdot \mathbf{t}_j'$, 
and
$([\mathbf{v}_{ij}' 
\otimes \mathbf{v}_{ij}' ] : \mathsf{M}_i ) \cdot\mathbf{n}_{ij}' 
= 2 A_i'(\boldsymbol{\rho}_i'-\mathbf{R}_i')\cdot \nabla(\nabla\cdot \mathbf{u}_i)
-\tfrac{1}{12}
  {\textstyle \sum_i}\overline{B}_{ij} \left(\left[\mathbf{t}_{j}'\otimes \mathbf{t}_{j}'\right] :\mathsf{M}_i\right)\cdot \mathbf{n}_{ij}'
\equiv 2A_i'(\boldsymbol{\rho}_i'-\mathbf{R}_i')\cdot [\nabla( \mathsf{I}: \mathsf{E})]_i 
-\tfrac{1}{12}
   {\textstyle \sum_i}\overline{B}_{ij} \left(\left[\mathbf{t}_{j}'\otimes \mathbf{t}_{j}'\right] :\mathsf{M}_i\right)\cdot \mathbf{n}_{ij}'$.
This gives
\begin{align*}
A_i&=A_i'\big[ 1+\mathsf{I}:\mathsf{E}_i + (\boldsymbol{\rho}_i'-\mathbf{R}_i')\cdot[\mathsf{I} :\nabla \mathsf{E}]_i \\
& +\tfrac{1}{2}(\boldsymbol{\rho}_i'-\mathbf{R}_i')\cdot[\nabla (\mathsf{I} : \mathsf{E})]_i
\big] \\
& +\tfrac{1}{24} \textstyle{\sum_j} \left( \overline{B}_{ij} \left[
    \mathbf{t}_{j}'\otimes \mathbf{t}_{j}' \right]: \mathsf{M}_i\right) \cdot\mathbf{n}_{ij}' \\
& -\tfrac{1}{24} \textstyle{\sum_j}
   \overline{B}_{ij}\left[ \mathbf{n}_{ij}'\cdot \left( \mathbf{t}_{j}' \cdot \mathsf{M}_i\right) \right] \cdot\mathbf{t}_{j}'.
\end{align*}
Then we note that
\begin{align*}
t_{j,p} & t_{j,q} n_{ij,r} M_{i,pqr}\\
&=-B_{ij}t_{j,p} t_{j,q} \epsilon_{i,rs}t_{j,s} (\partial_p E_{qr}+\partial_p W_{qr})_i \\
&=-B_{ij}t_{j,p} t_{j,q} \epsilon_{i,rs}t_{j,s} (\partial_p E_{qr}+\partial_p \varepsilon_{qr}\omega)_i \nonumber \\
=&\mp B_{ij}t_{j,p} t_{j,q} \varepsilon_{rs}t_{j,s} (\partial_p E_{qr}+\partial_p \varepsilon_{qr}\omega)_i \\
=&\mp B_{ij}t_{j,p} t_{j,q} t_{j,s} (\varepsilon_{rs}\partial_p E_{qr}-\delta_{qs} \partial_p \omega)_i \nonumber \\
=& \mp B_{ij}t_{j,p}  (t_{j,q} t_{j,s}\varepsilon_{rs}(\partial_p E_{qr})_i-t_{j}^2 (\partial_p \omega)_i) \\
= & \mp B_{ij}t_{j,p}  t_{j,q} t_{j,s}\varepsilon_{rs}(\partial_p E_{qr})_i-2t_{j}^2\mathbf{n}_{ij}\cdot \boldsymbol{\kappa}_i   \nonumber \\
=&\left(\mathbf{t}_j'\cdot\left[\mathbf{t}_j'\cdot(\nabla \mathsf{E})_i \right] \right)\cdot\mathbf{n}_{ij}' -2 t_j^{'2} \mathbf{n}_{ij}'\cdot \boldsymbol{\kappa}_i,
\end{align*}
taking $\boldsymbol{\epsilon}_i=\pm \boldsymbol{\varepsilon}$ and noting that $\mathbf{n}_{ij}\cdot\boldsymbol{\kappa}_i=\mp\tfrac{1}{2} B_{ij} t_{j,p}(\partial_p \omega)_i$ (with $\omega$ as defined in Appendix~\ref{sec:couple}). Similarly,
$t_{j,r} n_{ij,q} t_{j,p} M_{i,pqr}= \mp B_{ij}t_{j,r}  t_{j,p} t_{j,s}\varepsilon_{qs}(\partial_p E_{qr})_i+ 2t_{j}^2\mathbf{n}_{ij}\cdot \boldsymbol{\kappa}_i \\ = \left(\mathbf{n}_{ij}'\cdot \left[\mathbf{t}_j'\cdot (\nabla \mathsf{E})_i \right]\right)\cdot\mathbf{t}_j'   +2 t_j^{'2} \mathbf{n}_{ij}'\cdot \boldsymbol{\kappa}_i$.  Hence we can write
\begin{align}
A_i=&A_i'\Big[ 1+\mathsf{I}:\mathsf{E}_i + \tfrac{1}{2} (\boldsymbol{\rho}_i'-\mathbf{R}_i')\cdot (\nabla \mathsf{I}:\mathsf{E})_i  \nonumber \\
& + \mathsf{Y}_i \mathrel{\lower.03em\hbox{\rlap{$\cdot$}}{:}} (\nabla \mathsf{E})_i - \left[ {(6A_i')^{-1}}\textstyle{\sum_j} \overline{B}_{ij} (t_j')^2 \mathbf{n}_{ij}'\right] \cdot \boldsymbol{\kappa}_i \Big]
\label{eq:yi}
\end{align}
where $\mathsf{Y}_i$ has dimensions of length, with $Y_{i,pqr}=(\rho_{i,r}'-R_{i,r}')\delta_{pq}+(24 A_i')^{-1}{\textstyle \sum_j}\overline{B}_{ij} t_{j,p}' (t_{j,q}' n_{ij,r}'- n_{ij,q}'t_{j,r}')$.

\subsection{Energy and force potential}

Expanding the energy using (\ref{eq:rotf}), without the boundary component, gives
\begin{align*}
    \mathcal{U}&=\mathcal{U}'+{\textstyle \sum_{i,k}} \mathbf{f}_{ik}\cdot\delta\mathbf{r}_k \\
    & =\mathcal{U}'-{\textstyle \sum_{i,j,k}} B_{ij}A_{jk} (\boldsymbol{\epsilon}_i  \mathbf{h}_j)  \cdot\delta\mathbf{r}_k \\
    & =\mathcal{U}'-{\textstyle \sum_{i,j}} B_{ij} (\boldsymbol{\epsilon}_i  \mathbf{h}_j)  \cdot\delta\mathbf{t}_j \\
    & =\mathcal{U}'-{\textstyle \sum_{i,j}} B_{ij} (\boldsymbol{\epsilon}_i  \mathbf{h}_j)  \cdot \left[ \mathbf{t}_j'\cdot (\mathsf{E}_j+\mathsf{W}_j) \right] \\
    & =\mathcal{U}'-{\textstyle \sum_{i,j}} B_{ij} (\boldsymbol{\epsilon}_i  \mathbf{h}_j)  \cdot \big[ \mathbf{t}_j'\cdot (\mathsf{E}_i+\mathsf{W}_i \\
    & \qquad\qquad +\mathbf{v}_{ij}'\cdot(\nabla \mathsf{E})_i+\mathbf{v}_{ij}'\cdot (\nabla \mathsf{W})_i ) \big]
\end{align*}
using (\ref{eq:tjp}).  Using (\ref{eq:stresses}a), and taking $\boldsymbol{\epsilon}_i=\pm \boldsymbol{\varepsilon}$ this becomes
\begin{multline*}
    \mathcal{U} =\mathcal{U}' + {\textstyle\sum_i} A_i ( \boldsymbol{\sigma}^c : \mathsf{E}_i \pm \omega_i\, \{\mathrm{curl}^c\,\mathbf{h}\}_i ) \\ -{\textstyle \sum_{i,j}} B_{ij} (\boldsymbol{\epsilon}_i  \mathbf{h}_j)  \cdot \left[ \mathbf{t}_j'\cdot \left(\mathbf{v}_{ij}'\cdot(\nabla \mathsf{E})_i+\mathbf{v}_{ij}'\cdot (\nabla \mathsf{W})_i \right) \right].
\end{multline*}
For the chosen constitutive model, $\mathrm{curl}^c\,\mathbf{h}$ vanishes and rotation $\omega$ has no impact on energy.  The final term can be reformulated to give
\begin{align}
    \mathcal{U}& =\mathcal{U}' + {\textstyle\sum_i} A_i \boldsymbol{\sigma}^c : \mathsf{E}_i \nonumber \\
    & \qquad -{\textstyle \sum_{i,j}} B_{ij} (\boldsymbol{\epsilon}_i  \mathbf{h}_j)  \cdot \left[ \mathbf{t}_j'\cdot \left(\mathbf{v}_{ij}'\cdot(\nabla \mathsf{E})_i \right) \right] \nonumber \\
& \qquad    \mp {\textstyle \sum_{i,j}} B_{ij} (\mathbf{h}_j \cdot  \mathbf{t}_j')
    \mathbf{v}_{ij}'\cdot (2 \boldsymbol{\varepsilon} \boldsymbol{\kappa} )_i.
    \label{eq:conj}
\end{align}
 Gradients of strain and rotations will therefore generate hyperstresses conjugate to $\nabla \mathsf{E}$ and $\boldsymbol{\kappa}$, but that conjugate to $\boldsymbol{\kappa}$ is not equivalent to $\mathrm{CURL}^v\,\mathbf{h}$. 
 
\section{Computational implementation}
\label{sec:compimp}

The vertex model used in this paper was implemented as \texttt{VertexModel.jl} in the Julia programming language~\cite{bezansonJuliaFreshApproach2017,bezansonJuliaFastDynamic2012}, combining performance and readability. Figures were produced using the Julia \texttt{Makie.jl} plotting library~\cite{danischMakieJlFlexible2021}. All code used for this paper is available on GitHub \cite{Revell_VertexModel_jl_2022}.

We exploited incidence matrices $\mathsf{A}$ and $\mathsf{B}$ (following \cite{jensen2020}) to define a confluent epithelium. Systems were initialised as a symmetric arrangement of 7 cells with randomly allocated ages. $\mathsf{A}$ and $\mathsf{B}$ were updated following cell divisions, when a cell's age exceeds a specified cell cycle time, and neighbour exchanges (T1 transitions), when an edge length is shorter than a given threshold length. Both matrices have very low density and were stored as sparse arrays using the Julia \texttt{SparseArrays} library.  Vertex locations $\mathbf{r}$ were stored as a 1D vector of 2-component static array objects from the Julia \texttt{StaticArrays} library, optimising arithmetic operations on these position vectors. Data output and package dependencies were managed with the help of the \texttt{DrWatson.jl} package~\cite{datserisDrWatsonPerfectSidekick2020}.

After initialising all system arrays and packaging these into data structures for convenient passing between functions, positions $\mathbf{r}$ were updated using (\ref{eq:dynamics}), discretized with a fourth-order Runge-Kutta (RK) method. Spatial data and forces were recalculated at each RK substep, with topological matrices updated only once per full step. Monolayers were `grown' by simulating cell divisions for a specified number of cell cycle times, {\color{black}following random allocation of initial cell age within the first cycle.  The subsequent cell division time was substantially longer than the typical equilibration time.}  T1 transitions were implemented when edge lengths fell beneath a length 0.01, relative to a lengthscale in which the prescribed area is unity, as in (\ref{eq:energy}). Results were saved in \texttt{jld2} format and as an animated movie of the full simulation.

\end{appendices}



\bibliography{abbreviated}


\end{document}